\documentclass[10pt]{article}
\ifx\pdfoutput\undefined
\usepackage[dvips,bookmarks]{hyperref}
\else
\usepackage{hyperref}
\fi
\hypersetup{colorlinks=false,bookmarksopen,bookmarksnumbered,citecolor=blue,
   pdfstartview=FitH}

\usepackage[dvips]{graphicx}
\usepackage{latexsym}

\usepackage{amssymb,amsfonts,amsmath}
\usepackage{graphicx} 
\usepackage{indentfirst}

\usepackage{bbm}

\makeatletter \@addtoreset{equation}{section} \makeatother

\usepackage[usenames]{color}

\newcommand {\cC}{{\cal C}}
\newcommand {\cD}{{\cal D}}

\newcommand {\cF}{{\cal F}}
\newcommand {\cG}{{\cal G}}
\newcommand {\cH}{{\cal H}}

\newcommand {\cK}{{\cal K}}
\newcommand {\cL}{{\cal L}}
\newcommand {\cM}{{\cal M}}
\newcommand {\cN}{{\cal N}}
\newcommand {\cO}{{\cal O}}

\newcommand {\cR}{{\cal R}}
\newcommand {\cS}{{\cal S}}
\newcommand {\cT}{{\cal T}}
\newcommand {\cU}{{\cal U}}
\newcommand {\cV}{{\cal V}}
\newcommand {\cW}{{\cal W}}

\newcommand{\bH}{{\bf H}}

\def\a{\alpha}
\def \bi{\bibitem}

\def\b{\beta}

\def\d{\delta}
\def\e{\epsilon}
\def\f{\phi}
\def\g{\gamma}
\def\G{\Gamma}

\def\j{\psi}
\def\k{\kappa}
\def\l{\lambda}
\def\m{\mu}
\def\n{\nu}
\def\o{\omega}
\def\p{\pi}
\def\q{\theta}

\def\s{\sigma}
\def\t{\tau}

\def\x{\xi}
\def\z{\zeta}
\def\D{\Delta}
\def\F{\Phi}
\def\J{\Psi}
\def\L{\Lambda}
\def\O{\Omega}
\def\P{\Pi}
\def\Q{\Theta}
\def\S{\Sigma}
\def\U{\Upsilon}
\def\X{\Xi}
\def\tr{{\rm tr}}
\def\rd{{\rm d}}
\def\ri{{\rm i}}

\newcommand{\ad}{{\dot{\alpha}}}                           
\newcommand{\bd}{{\dot{\beta}}}                            
\newcommand{\ve}{\varepsilon}                            

\newcommand{\pa}{\partial}                           
\newcommand{\hf}{\frac12}

\newcommand{\vf}{\varphi}

\newcommand{\be}{\begin{equation}}
\newcommand{\ee}{\end{equation}}
\newcommand{\bea}{\begin{eqnarray}}
\newcommand{\eea}{\end{eqnarray}}
\newcommand{\non}{\nonumber}
\newcommand{\1}{\underline{1}}
\newcommand{\2}{\underline{2}}

\def\dt#1{{\buildrel {\hbox{\LARGE .}} \over {#1}}}    

\newcommand{\bm}[1]{\mbox{\boldmath$#1$}}

\def\double #1{#1{\hbox{\kern-2pt $#1$}}}

\begin{document}

\begin{center}
{\large \bf  
LECTURES ON NONLINEAR SIGMA-MODELS\\
IN PROJECTIVE SUPERSPACE}\footnote{Invited lectures  presented at  the 30th Winter School 
{\it GEOMETRY AND PHYSICS},  Srni, Czech Republic, 
16-- 23 January, 2010.}
\\ 
\end{center}

\begin{center}

{\bf
Sergei M. Kuzenko
} \\
\vspace{5mm}

\footnotesize{
{\it School of Physics M013, The University of Western Australia\\
35 Stirling Highway, Crawley W.A. 6009, Australia}\\
{\tt kuzenko@cyllene.uwa.edu.au}
}  
~\\
\vspace{2mm}

\end{center}
\vspace{5mm}

\begin{abstract}
\baselineskip=14pt
$\cN = 2$ supersymmetry in four space-time dimensions  is intimately related to hyperk\"ahler and quaternionic 
K\"ahler geometries. On one hand, the target spaces for rigid supersymmetric sigma-models are necessarily 
hyperk\"ahler manifolds.  On the other hand, when coupled to $\cN = 2$ supergravity, the sigma-model target 
spaces must be quaternionic K\"ahler. It is known that such manifolds of restricted holonomy are difficult  to generate explicitly. Projective superspace is a field-theoretic approach to construct general $\cN = 2$ supersymmetric nonlinear sigma-models, and hence to generate  new hyperk\"ahler and  quaternionic K\"ahler metrics. Intended for a mixed audience consisting of both physicists and mathematicians, these lectures provide a pedagogical introduction to the projective-superspace approach.  
\end{abstract}
\vspace{0.3cm}


\renewcommand{\thefootnote}{\arabic{footnote}}
\setcounter{footnote}{0}

\tableofcontents{}
\bigskip\hrule

\newpage

\begin{flushright}\includegraphics[width=.3\textwidth]{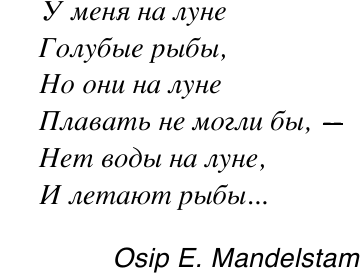}\end{flushright}


\section{Introduction}
\setcounter{equation}{0}
The concept of supersymmetry in four space-time dimensions 
was introduced in theoretical  physics in the early 1970s \cite{GL,VA,WZ}.
It is a symmetry between bosons and fermions in relativistic theories (field theory, string theory, etc.).
The discovery of supersymmetry led, in a short period of time,  to the appearance of new research directions 
in high-energy physics, due to remarkable properties of supersymmetric theories, including the following:
\begin{itemize} 

\item Supersymmetry has  nontrivial manifestations at the quantum level;

\item Local supersymmetry implies gravity (supergravity \cite{N=1SUGRA});

\item One version of local supersymmetry ($\cN=2$ supergravity \cite{FvN}) fulfills Einstein's dream of 
unifying gravity and electromagnetism;

\item String theory requires supersymmetry.

\end{itemize} 
These studies mostly involved the physics community. However, in the late 1970s and early 1980s
supersymmetry met complex geometry. 

The year 1979,
 the Einstein centennial year, was special for physics  and geometry. 
On the physics side, a work of Zumino \cite{Zumino} uncovered 
an intimate connection between supersymmetry and complex geometry. 
On the geometry side, Calabi \cite{Calabi2} introduced the concept of hyperk\"ahler geometry.
The fact that the two discoveries took place in the same year, was just a coincidence. 
However, what followed in the next 30 years
was a remarkably fruitful interaction between supersymmetry and hyperk\"ahler geometry. 
An example of this  is an influential  paper
by  Hitchin, Karlhede, Lindstr\"om and Ro\v cek \cite{HitchinKLR}.
The present  lectures will give an overview of some of these developments.

Nontrivial evidence for the existence of connections between supersymmetry and complex geometry
comes from the consideration of supersymmetric nonlinear sigma-models. There are three relevant classic results:
\begin{itemize} 

\item K\"ahler manifolds are target spaces for rigid supersymmetric 
sigma-models with four supercharges  (${\rm D}\leq 4$) \cite{Zumino}.
In four dimensions, D = 4, such sigma-models possess $\cN=1$ supersymmetry;

\item Hyperk\"ahler manifolds are target spaces for rigid supersymmetric 
sigma-models with eight supercharges  (${\rm D}\leq 6$) \cite{A-GF}.
In four dimensions, such sigma-models possess $\cN=2$ supersymmetry;

\item Quaternionic K\"ahler  manifolds are target spaces for locally supersymmetric 
sigma-models with eight supercharges (${\rm D}\leq6$) \cite{BW}. 

\end{itemize} 
Supersymmetric sigma-models generalize ordinary bosonic ones.
It is pertinent here to  recall that a bosonic nonlinear sigma-model is a field theory over a space-time 
$\mathbb X$ in which the fields
take values in a $d$-dimensional Riemannian manifold $(\cM^d, g)$ (known as {\it target space}). 
If $\mathbb X$  is four-dimensional  Minkowski space,  ${\mathbb M}^4$, 
the sigma-model action is 
\bea
S=-\hf \int {\rm d}^4 x \, g_{\m \n} (\vf ) \pa^a  \vf^\m \pa_a \vf^\n ~,
\eea
where $\vf^\m (x)$ are {scalar fields} on ${\mathbb M}^4$  
and {local coordinates} on $\cM^d$ (more precisely, the field $\vf (x)$ takes its values in  $\cM^d$).

Unlike K\"ahler metrics, the hyperk\"ahler and quaternionic K\"ahler metrics are rather difficult
to generate explicitly \cite{Besse}.  In this regard, it turns out that 
the sigma-model results of \cite{A-GF,BW} 
have an important implication that was not immediately recognized and appreciated. The idea is that  
off-shell $\cN=2$ supersymmetry, {\it provided its power is properly elaborated},  
is a device to generate hyperk\"ahler and quaternionic K\"ahler structures
\cite{LR,KLR,GIOS86,HitchinKLR}. 
More precisely, suppose it is possible to develop a formalism for constructing 
$\cN=2$ rigid supersymmetric sigma-models generated by a Lagrangian of reasonably 
general functional form (say, an arbitrary real analytic function of several variables). 
Then, for any choice of the Lagrangian, the target space metric must be hyperk\"ahler. 
Any deformation of the Lagrangian will lead to a new $\cN=2$ sigma-model, 
and hence to a new hyperk\"ahler metric.

It appears that the only way to make the above  idea work is to develop $\cN=2$ {\it superspace techniques} 
for constructing general $\cN=2$ supersymmetric sigma-models. Indeed, superspace is known 
to provide unique opportunities to engineer supersymmetric theories.
 Two fully-fledged $\cN=2$ superspace approaches
have been developed: (i) harmonic superspace
 \cite{GIKOS,GIOS}; and (ii) projective superspace   
 \cite{KLR,GHR,LR-projective1,LR-projective2,G-RRWLvU}. 
The former is  more general\footnote{Off-shell projective multiplets and their couplings 
can be obtained from those emerging within the harmonic-superspace approach via   a singular truncation 
of multiplets \cite{K98} or, equivalently, by integrating out some auxiliary degrees of freedom \cite{JS}.}
\cite{K98,JS}; it is also  powerful 
in the context of quantum  $\cN=2$ super Yang-Mills theories. 
However, it is the latter  approach which is ideally designed for sigma-model
constructions. These notes provide a pedagogical introduction to the projective-superspace approach.  
 
It should be noted that the problem of generating  quaternionic K\"ahler metrics can be reduced to 
that of hyperk\"ahler ones. 
There exists  a remarkable one-to-one correspondence between $4n$-dimensional quaternionic K\"ahler 
manifolds and $4(n+1)$-dimensional hyperk\"ahler spaces possessing a homothetic conformal Killing 
vector, and hence an isometric action of SU(2) rotating the complex structures
\cite{Swann} (see also \cite{Galicki}).
Such hyperk\"ahler spaces are called  {\it Swann bundles} in the  mathematics literature
\cite{BoyG},  and {\it hyperk\"ahler cones}  in the physics literature \cite{deWRV}.
Hyperk\"ahler cones
are target spaces for $\cN=2$ rigid superconformal sigma-models \cite{deWRV,SezginT}.
Therefore, it is sufficient to develop techniques to generate arbitrary 
 $\cN=2$ rigid supersymmetric nonlinear  sigma-models, and hence
hyperk\"ahler metrics.

These notes are organized as follows.  
In order to make our presentation reasonably self-contained and accessible to mathematicians, 
 two introductory sections are included.
Section 2 is  devoted to  algebraic aspects of supersymmetry (the $\cN$-extended 
super-Poincar\'e group, its algebra, superspace), while section 3 presents
 elements of field theory in superspace. 
Section 4 describes the formulations of $\cN=1$ and $\cN=2$ supersymmetric nonlinear sigma-models
in terms of $\cN=1$ chiral superfields. 
In section 5, we introduce an extension of the $\cN=2$ conventional superspace by auxiliary bosonic 
directions, 
\bea
{\mathbb M}^{4|8} \quad \longrightarrow \quad {\mathbb M}^{4|8}\times {\mathbb C}P^1
\equiv {\mathbb M}^{4|8}\times S^2~,
\eea
and  give a brief introduction to  the harmonic and projective superspace approaches.
Off-shell projective supermultiplets and related constructions are discussed in section 6.
In section 7, we present the most general $\cN=2$ off-shell supersymmetric nonlinear sigma-models
in projective superspace, and then review two versions of the Legendre transform construction: 
the generalized and linear ones. As an application of the methods developed, 
in sections 8 and 9 we review various aspects of the $\cN = 2$ supersymmetric sigma-models on cotangent bundles 
of K\"ahler manifolds. Section 10 includes comments on the topics not covered.
This paper is concluded with two technical appendices. 
Appendix A is devoted to the $\cN$-extended superconformal group in four space-time dimensions. 
Appendix B contains essential information about canonical coordinates for K\"ahler manifolds.

Our notation and two-component spinor conventions correspond to those used in two textbooks
\cite{WB,BK}. In particular, the Minkowski metric is chosen to be $\eta_{mn}= {\rm diag} \,(-1, +1, +1, +1)$.
A brief summary of the two-component (iso)spinor conventions is given in Appendix C.

\section{Algebraic aspects of supersymmetry}
\setcounter{equation}{0}

In our presentation of the $\cN$-extended super-Poincar\'e group and superspace, 
we follow the 1973 paper by Akulov and Volkov \cite{AV} in which these concepts 
were introduced for the first time.\footnote{The Akulov-Volkov paper \cite{AV}
was submitted to the journal {\it Theoretical  and Mathematical Physics} on 8 January 1973, 
and published in January 1974. It remains largely unknown, probably because
it was published in a Russian journal. 
The concepts of  the $\cN$-extended super-Poincar\'e group and superspace have been discussed 
in many books and reviews, however the pioneering approach of \cite{AV}
is still one of the best.}

\subsection{Matrix realization of the Poincar\'e group}

Denote by $ {\mathfrak P}(4)$ the universal covering group of 
the restricted Poincar\'e group $ {\rm ISO}_0(3,1)$. The principle of
relativistic invariance states 
that  $ {\mathfrak P}(4)$ must be a subgroup of the symmetry group 
of any quantum field theory.

Traditionally,   ${\mathfrak P}(4)$ is realized as the group of linear inhomogeneous transformations 
on the space of $2\times 2$ Hermitian matrices (with $ \vec \s$ being the Pauli matrices)
\bea
{\bm x}: = x^m \s_m ={\bm x}^\dagger =( x_{\a \dt \b})~, 
\qquad { \s_m = ({\mathbbm 1}_2, \vec{\s} )}~, 
\qquad x^m \in {\mathbb R}^4
\eea
defined to act as follows:
\bea
{\bm x} ~\to ~{\bm x}' =  x'^m \s_m =M{\bm x}M^\dagger + {\bm b} ~, 
\qquad {\bm b} = b^m \s_m ~,
\eea
with
\bea
M = (M_\a{}^\b ) \in {\rm SL}(2,{\mathbb C})~,  \qquad b^m \in {\mathbb R}^4~.
\eea
Here $M^\dagger := {\bar M}^{\rm T}$ denotes the Hermitian conjugate of $M$,
and $\bar M =({\bar M}_{\dt \a}{}^{\dt \b} )$ the complex conjugate of $M$, 
with ${\bar M}_{\dt \a}{}^{\dt \b} := \overline{ M_\a{}^\b}$. 

The above realization of ${\mathfrak P}(4)$  admits a useful  equivalent form,
as the group of linear inhomogeneous transformations 
  on the space of $2\times 2$ Hermitian matrices 
\bea
\tilde{\bm x}: = x^m \tilde{\s}_m =\tilde{\bm x}{}^\dagger =({ x^{\dt \a \b}})~, 
\qquad { \tilde{\s}_m = ({\mathbbm 1}_2, -\vec{\s} )}~, 
\qquad x^m \in {\mathbb R}^4
\eea
defined to act as follows:
\bea
\tilde{\bm x} ~\to ~\tilde{\bm x}' =  x'^m \tilde{\s}_m =(M^{-1})^\dagger \tilde{\bm x}M^{-1} + \tilde{\bm b} ~,
\qquad \tilde{\bm b} = b^m \tilde{\s}_m ~.
\label{ISL(2,C)}
\eea
The matrices $ ({\s}_m)_{\a\dt \b}  $ and 
$ (\tilde{\s}_m )^{\dt \a \b}$ turn out to be  invariant tensors of the restricted Lorentz group $ {\rm SO}_0(3,1)$, 
and they transform into each other under space reflection $x^m =(x^0, \vec{x}) \to x'^m:=(x^0, -\vec{x})$.

${}$For our subsequent consideration, it is advantageous to
realize ${\mathfrak P}(4)$ as a subgroup of  the group ${\rm SU}(2,2)$,
which is  a  4--1 covering of the conformal group in four space-time dimensions,   
consisting of all block triangular matrices of the form:
\bea
&& { (M, {b})} := \left(
\begin{array}{c | c}
  M  ~& ~ 0   \\
\hline
-{\rm i}\,\tilde{\bm b} \,M 
\phantom{\Big(}
& ~(M^{-1})^\dagger\\
\end{array}
\right) = { (\mathbbm{1}_2, {b}) \, (M,0 )}~,
\label{(N,b)} \\
&&M  \in  {\rm SL}(2,{\mathbb C})~, \qquad 
\tilde{\bm b} := b^m \tilde{\s}_m =\tilde{\bm b}^\dagger ~, 
\qquad b^m \in {\mathbb R}^4~.
\non
\eea

It is well known that {Minkowski space} $\mathbb{M}^4\equiv {\mathbb R}^{3,1}$
is a homogeneous space of the Poincar\'e group, 
and can be identified with the coset space $ {\rm ISO}_0(3,1)/{\rm SO}_0(3,1) $. 
However, it can  equivalently be realized as the coset space 
\bea
{\mathbb M}^4= {\mathfrak P}(4)/ {\rm SL}(2,{\mathbb C})~.
\eea 
Its points are naturally  parametrized by the Cartesian coordinates $x^m \in {\mathbb R}^4$ 
corresponding to the coset representative:
\bea
(\mathbbm{1}_2, x) = 
\left(
\begin{array}{r | c}
  \mathbbm{1}_2  ~& ~ 0   \\
\hline
-{\rm i}\,\tilde{\bm x}  ~& ~\mathbbm{1}_2\\
\end{array}
\right) = \exp  \left(
\begin{array}{r | c}
 0 ~& ~ 0   \\
\hline
-{\rm i}\,\tilde{\bm x}  ~& ~ 0 \\
\end{array}
\right)~ .
\label{coset-represent}
\eea
${}$From here one can read off 
the action of ${\mathfrak P}(4)$ on $\mathbb{M}^4$: 
\bea
(M,b) \,(\mathbbm{1}_2, x) = (\mathbbm{1}_2, x') \,(M,0) \quad
\Longleftrightarrow \quad 
x'^m = \big( \L (M) \big)^m{}_n \,x^n+b^m~.
\label{PoincareTrans}
\eea
Here $ \L \!: {\rm SL}(2,{\mathbb C}) \to {\rm SO}_0(3,1)$ 
is the {doubly covering homomorphism} defined by
\bea
\big( \L (M) \big)^m{}_n = -\hf \tr \big( \tilde{\s}{}^m M \s_n M^\dagger\big)~.
\eea 
The right-hand side of (\ref{PoincareTrans}) 
coincides with  the standard action of ${\rm ISO}_0(3,1)$ on Minkowski space.

\subsection{Matrix realization of the super-Poincar\'e group}
Supersymmetry is the only consistent and  nontrivial extension of the Poincar\'e  symmetry 
that is compatible  with the principles of quantum field theory \cite{HLS}.

Denote by  ${\mathfrak P}{(4|\cN )}$ the 
$\cN$-extended super-Poincar\'e group. It can be realized as 
a subgroup of $ {\rm SU}(2,2|\cN)$, the $\cN$-extended superconformal group
(see Appendix A for its definition).
Any element $g \in {\mathfrak  P}{(4 | \cN )}$ is a $(4+\cN)\times (4+\cN)$ 
{supermatrix} of the form:
\begin{subequations}
\bea
g &=& {s(b, {\bm \ve}) \,h (M) } ~, \qquad {\bm \ve} := (\e^\a_i , {\bar \e}_{\dt \a }^i)~, 
\qquad i=1,\dots,\cN
\label{SP1} \\
{ s(b,{\bm \ve} )} &:= &
\left(
\begin{array}{c | c ||c}
  \mathbbm{1}_2  ~& ~ 0&~{ 0}   \\
\hline
-{\rm i}\,\tilde{\bm b}_{(+)}  ~& ~\mathbbm{1}_2 &~ {  2{\bar \e}} \\
\hline
\hline
{ 2\e} ~& ~{ 0}&~\mathbbm{1}_\cN
\end{array}
\right)
=
\left(
\begin{array}{c | c ||c}
\d_\a{}^\b  ~& ~ 0&~{ 0}   \\
\hline
-{\rm i}\,b^{{\dt \a} \b}_{(+)}  ~& ~\d^{\dt \a}{}_{\dt \b} &~ { 2{\bar \e}^{{\dt \a} j}}
\label{SP-g} \\
\hline
\hline
{2\e_i{}^\b} ~& ~{ 0}&~\d_i{}^j
\end{array}
\right) ~, 
\label{SP2}  \\
{ h(M)} &:=& 
\left(
\begin{array}{c | c ||c}
 M  ~& ~ 0&~{ 0}   \\
\hline
0  ~& ~(M^{-1})^\dagger&~ { 0} \\
\hline
\hline
{ 0} ~& ~{ 0}&~\mathbbm{1}_\cN
\end{array}
\right) 
= \left(
\begin{array}{c | c| |c}
M_\a{}^\b  ~& ~ 0&~{ 0}   \\
\hline
0  ~& ({\bar M}^{-1})_{\dt \b}{}^{\dt \a}  &~ {  0} \\
\hline
\hline
{  0} ~& ~{  0}&~\d_i{}^j
\end{array}
\right) ~, 
\label{SP3} 
\eea
\end{subequations}
where $M \in {\rm SL}(2,{\mathbb C})$ and  
\bea
 b^m_{(\pm )}:= b^m \pm {\rm i} \,\e_i \s^m {\bar \e}^i 
=b^m \pm {\rm i}\, \e_i^\a (\s^m)_{\a \ad} {\bar \e}^{{\dt \a} i}  ~, 
\qquad \overline{b^m}=b^m~.
\label{+-}
\eea
The group element $s(b, {\bm \ve})$ is generated by 
four {\it commuting} (or bosonic) real parameters $b^m$, 
$2\cN$ {\it anti-commuting} (or fermionic)
complex parameters $\e^\a_i$  and their complex conjugates 
$ {\bar \e}^{{\dt \a} i} $, 
$  {\bar \e}^{{\dt \a} i} :=
\overline{\, \e^\a_i \,}$.
In supersymmetric quantum field theory, the various elements of ${\mathfrak P}{(4|\cN )}$
correspond to several different symmetries, specifically:
$h(M)$ describes a {Lorentz transformation}, 
$s(b, 0)$ a {space-time translation}, and 
$s(0,{\bm  \e})$  a {supersymmetry transformation}.

It is easy to check that the set of supermatrices ${\mathfrak P}{(4|\cN )}$ introduced
is a group. This follows from the  easily verified identities:
\begin{subequations}
\bea
 s(b, {\bm \ve}) s(c, {\bm \eta} ) &=& s(d, {\bm \ve} +{\bm \eta}) ~,
\label{group1} \\
 h(M) s(b, {\bm \ve}) h(M^{-1}) &=& s \big(\L(M) b, 
 \hat{\bm \ve} \big)~,
\label{group2}
\eea 
\end{subequations}
where
we have defined 
\bea
 d^m :=b^m +c^m +{\rm i} \Big(\eta_i\s^m {\bar \e}^i -\e_i \s^m {\bar \eta}^i \Big)~, 
 \qquad 
 \hat{\bm \ve}:= {\bm \ve} 
  \left(
\begin{array}{c | c}
  M^{-1}  ~&  ~0   \\
\hline
0
\phantom{\Big(}
& ~M^\dagger\\
\end{array}
\right) ~.~~~
\eea

By definition, {$\cN$-extended Minkowski superspace} is  the homogeneous space
\bea
{\mathbb M}^{4|4\cN} 
={\mathfrak P}(4|\cN)\big/ {\rm SL}(2,{\mathbb C}) 
~,
\eea
where ${\rm SL}(2,{\mathbb C})$ is now identified with the set of all matrices 
$h(M)$.
The points of ${\mathbb M}^{4|4\cN}$
can be parametrized by the variables 
\bea
 z^M = (x^m, \q^\a_i , {\bar \q}_{\dt \a}^i ) \equiv (x,  \Q )
\eea
which correspond to the following coset representative:
\bea
{s(z):=s(x,\Q )} 
&=& \left(
\begin{array}{c | c ||c}
  \mathbbm{1}_2  ~& ~ 0~&~0   \\
\hline
-{\rm i}\,\tilde{\bm x}_{(+)}  ~& ~\mathbbm{1}_2 ~&~ 2{\bar \q} \\
\hline
\hline
2\q ~& ~0~&~\mathbbm{1}_\cN
\end{array}
\right)  
= \  \exp 
 \left(
\begin{array}{c | c ||c}
 0  ~& ~ 0~&~0   \\
\hline
-{\rm i}\,\tilde{\bm x}  ~& ~0 ~&~ 2{\bar \q} \\
\hline
\hline
2\q & ~0~&~0
\end{array}
\right)  ~.~~~
\label{s(z)}
\eea
The action of ${\mathfrak P}(4|\cN)$ on ${\mathbb M}^{4|4\cN}  $ is naturally defined by 
\bea
g= s(b, {\bm \ve}) \,h (M): \quad  s(z) ~ \to ~s(z') := 
s(b, {\bm \ve}) h (M) s(z) h(M^{-1})~.~~~
\label{Gaction}
\eea
Using this definition allows one to read off 
a {\it Poincar\'e transformation} associated with   $g= s(b, 0) \,h (M)$ 
\bea
x'^m = \big( \L (M) \big)^m{}_n \,x^n+b^m~, 
\qquad 
\q'^\a_i = \q^\b_i (M^{-1})_\b{}^\a~,
\eea
as well as a
{\it supersymmetry transformation}  corresponding to $g = s(0, {\bm \ve})$ 
\bea
x'^m = x^m + {\rm i} \big(\q_i\s^m {\bar \e}{}^i 
-\e_i \s^m {\bar \q}^i \big)~, \qquad \q'^\a_i = \q^\a_i + \e^\a_i~.
\eea

\subsection{The super-Poincar\'e algebra}
We can represent group elements of ${\mathfrak P} (4|\cN )$ in an exponential form:
\begin{subequations}
\bea
s(b, {\bm \ve}) &=& \exp {\rm i} \,\Big\{ -b^m { P_m} + \e^\a_i \, {  Q^i_\a} 
+ {\bar \e}^i_{\dt \a} \, { {\bar Q}^{\dt \a}_i } \Big\} ~,  \\
h( {\rm e}^{\hf \o^{mn} \s_{mn} }) &=& 
\left(
\begin{array}{c | c ||c}
  {\rm e}^{\hf \o^{mn} \s_{mn} } & ~ 0&~{0}   \\
\hline
0  & ~ {\rm e}^{\hf \o^{mn} \widetilde{\s}_{mn} } &~ { 0} \\
\hline
\hline
{ 0} & ~{ 0}&~\mathbbm{1}_\cN
\end{array}
\right) = \exp \Big\{\frac{\rm i}{2} \,\o^{mn}  J_{mn}\Big\}~,~~~~~~~~
\label{SP-lor}
\eea
\end{subequations}
where  $\o^{mn}=-\o^{nm}$ are real parameters, and 
\bea
\s_{mn}:= -\frac{1}{4}\Big( \s_m \tilde{\s}_n -  \s_n \tilde{\s}_m \Big)~,\qquad
\widetilde{\s}_{mn}:= -\frac{1}{4}\Big( \tilde{\s}_m {\s}_n -  \tilde{\s}_n {\s}_m \Big)~.
\eea
Here ${ P_m}$, $ J_{mn}$, 
${ Q^i_\a} $ and ${ {\bar Q}^\ad_i }$ are 
the generators of the Lie superalgebra $ {\mathfrak p} (4|\cN )$ of ${\mathfrak P} (4|\cN )$.
In field-theoretic representations of $ {\mathfrak p} (4|\cN )$, 
 $P_m =(-E, \vec{P})$  is identified with  the energy-momentum 4-vector, $J_{mn}$ the Lorentz generators, 
 and $Q_\a^i $ and ${\bar Q}^{\dt \a}_i$ the supersymmetry generators.

Making use of eq. (\ref{group1}), one can derive 
 the (anti-)commutation relations:
\begin{subequations}
\bea
\big[P_m , P_n \big]& =& 0~,  \label{2.23a} \\
 \big [P_m, Q_\a^i \big] &=&  \big[P_m , {\bar Q}_{\dt \a i}  \big] =0~,   \label{2.23b}\\
\{Q^i_{\a} \, , \, Q^j_{ \b} \} &=& 
\{{\bar Q}_{\dt \a i} \, , \, {\bar Q}_{\dt  \b j} \} = 0~,  
 \label{2.23c}  \\
\{ Q_\a^i \, , \, {\bar Q}_{\dt  \b j} \} &=& 2 \d^i_j\,(\s_m)_{\a \dt \b} \,P^m~. 
 \label{2.23d}
\eea
\end{subequations}
In conjunction with commutation relations involving the Lorentz generators, 
which can be readily derived with the aid of (\ref{SP-lor}),
the above (anti-)commutation relations constitute the $\cN$-extended 
super-Poincar\'e algebra. The $\cN=1$ super-Poincar\'e algebra
was discovered in 1971 by Golfand and Likhtman 
\cite{GL}.

\subsection{Adding the R-symmetry group}
The super-Poincar\'e algebra ${\mathfrak p} (4|\cN )$ has  a nontrivial group of  
outer automorphisms that is isomorphic to ${\rm U}(\cN)$ and is known as the $R$-symmetry group.
The {$\cN$-extended super-Poincar\'e group} ${\mathfrak P}{(4|\cN )}$ 
can be generalized to include  the $R$-symmetry group.
The resulting supergroup is denoted  $ {\mathfrak P}_{\rm A}{(4|\cN )}$.
Any element $g \in {\mathfrak  P}_{\rm A} {(4 | \cN )}$ is a $(4+\cN)\times (4+\cN)$ 
supermatrix of the form \cite{AV}:
\begin{subequations}
\bea
g &=& { s(b, {\bm \ve}) \,h (M, U) } ~, 
\qquad {\bm \ve} := (\e^\a_i , {\bar \e}_{\dt \a }^i)~, 
\qquad i=1,\dots,\cN ~~~~
 \\
{ s(b,{\bm \ve} )} &:= &
\left(
\begin{array}{c | c ||c}
  \mathbbm{1}_2  ~& ~ 0&~{ 0}   \\
\hline
-{\rm i}\,\tilde{\bm b}_{(+)}  ~& ~\mathbbm{1}_2 &~ { 2{\bar \e}} \\
\hline
\hline
{ 2\e} ~& ~{ 0}&~\mathbbm{1}_\cN
\end{array}
\right) ~,
\\
{ h(M, U)} &:=& 
\left(
\begin{array}{c | c ||c}
 M  ~& ~ 0&~{ 0}   \\
\hline
0  ~& ~(M^{-1})^\dagger&~ { 0} \\
\hline
\hline
{  0} ~& ~{  0}&~U
\end{array}
\right) ~,
\qquad U =(U_i{}^j) \in {\rm U}(\cN)~.
\eea
\end{subequations}

\noindent
{$\cN$-extended Minkowski superspace} is  the homogeneous space
\bea
{\mathbb M}^{4|4\cN} 
={\mathfrak P}_{\rm A}(4|\cN)\big/ {\rm SL}(2,{\mathbb C}) \times {\rm U}(\cN)
~.
\eea

In the case $\cN>1$, the super-Poincar\'e algebra can be further generalized to include
central charges \cite{HLS}. Such a generalization was not considered in \cite{AV}.

\section{Field theory in superspace}
\setcounter{equation}{0}

This section is a mini-introduction to supersymmetric field theory.
It contains only those concepts and results that we consider absolutely essential 
for the subsequent discussion of supersymmetric nonlinear sigma-models. 
Comprehensive reviews of 
 supersymmetric field theory can be found, e.g., in the textbooks \cite{WB,BK,GGRS}.

\subsection{A brief review of the coset construction}
Here we succinctly review   the salient points of  Cartan's coset construction. From 
the point of view of a theoretical physicist, this is a procedure to develop a field 
theory on a homogeneous space $X= \{x\}$ of a Lie group $G$. 
The  homogeneous space can always be realized as a left coset space
\bea
X = G/H ~, 
\eea
for some closed subgroup $H$ of $G$. 
We denote by $\p$ the natural projection, $ \p \!: G \to G/H$, defined by 
$\p ( g) = gH$, for any $g \in G$.

${}$For simplicity, we assume the  existence\footnote{Quite often, no global cross-section exists, 
and then one has to restrict
the consideration to local coordinate
charts. For example, this happens if $X =S^2 ={\rm SU(2)/U(1)}$ and  $G= {\rm SU(2)}$.
However, in some cases of interest, one can construct such a  global cross-section. 
This is indeed the case if $X=  {\mathbb M}^{4|4\cN}$ and 
$G $ coincides with  ${\mathfrak P}{(4|\cN )}$ or  $ {\mathfrak P}_{\rm A}{(4|\cN )}$.} 
of  a global {\it cross-section} (also known as {\it coset representative})
$s(x) \!: X \to G$ {such that }
\bea
\p\circ s = {\rm id} 
\qquad \Longleftrightarrow \qquad \p \big(s(x)\big) =x ~,\quad \forall x \in X~.
\eea
We then have the following 
unique decomposition in the Lie group $G$: for any group element $ g \in G $ 
there exist  unique  $x \in X$ and $h \in H$ such that
\bea
{ g = s(x) \, h}~.
\eea
Now, the fact that $G$ acts on $X =G/H$ can be expressed as follows:
\bea
{ g\, s(x) = s (g\cdot x) \, {\bm h} (g,x)} \equiv  { s (x')} \, {\bm h} (g,x)~, 
\qquad
{\bm h} (g,x) \in H
\eea
where ${\bm h} (g,x) $ obeys the property (see, e.g.,  \cite{Kirillov})
\bea
{\bm h} (g_1 g_2,x) = {\bm h} (g_1, g_2 x) \,{\bm h} (g_2,x) ~.
\eea

Let $R$ be a finite-dimensional representation of $H$ on a vector space $\cV$. 
We then can define a representation $T$ of $G$ acting on a linear space of fields $\vf(x)$ over $X$ 
with  values in $\cV$, 
$\vf: X \to \cV$,
by the rule:
\bea
 \Big[ T(g) \vf \Big] (g\cdot x)} \equiv {  \vf'(x') 
={  R \Big( {\bm h} (g,x) \Big) \vf(x) }~.
\label{IndRep1}
\eea
The representation $T$ is called {\it induced} (more precisely, the representation of $G$ induced 
by the representation $R$ of the subgroup $H$), see, e.g. \cite{Kirillov} for more details.
The notion of induced representation is indispensable to quantum field theory. 
The point is that all relativistic fields we deal with in physics, are examples of 
this construction.

The notion of induced representation can be reformulated in a way that requires no use of 
${\bm h} (g,x) $ \cite{Kirillov}. Consider a linear space of $\cV$-valued functions on $G$, $\f (g)$, 
such that  $\f (g \,h^{-1}) = R (h) \f(g)$, for arbitrary $g \in G$ and $h \in H$. On this space, we can define a representation $T$ of $G$ by 
\bea
\Big[ T(g) \f \Big] (g_0) = \f (g^{-1} g_0)~,
\eea
which can be seen to be equivalent to the induced one.
Indeed, the  construction under consideration 
reduces to that considered above by introducing $\vf(x) := \f \big(s(x)\big) $.

Our next task is to learn how to differentiate fields $\vf(x) $ over $X$
in a $G$-covariant way.
Denote by $ \cG$ and $ \cH$ the Lie algebras of $G$ and $H$, respectively. 
Suppose that there exists   a complement $ \cK$  of $\cH$ in $\cG$ which is invariant under
the adjoint representation of $H$ on the Lie algebra $\cG$. Thus we have
\bea
\cG = \cK \oplus \cH
 ~, \qquad [\cH, \cH] \in \cH~, \qquad [\cH, \cK] \in \cK~.
\eea
Let $ \{\cT_\a \}$ be a basis of $\cK$, and $ \{\cT_i \}$  a basis for $\cH$. 
Introduce the left-invariant {\it Maurer-Cartan one-form}:
\begin{subequations}
\bea
 s^{-1} {\rm d} s &=& E+ \O~, \\
 E &=& {\rm d} x^\m E_\m{}^\a (x) \cT_\a \equiv E^\a \cT_\a~,\\
 \O &=& {\rm d}x^\m \,\O_\m{}^i(x) \cT_i \equiv E^\a\, \O_\a{}^i \cT_i~.
\eea
\end{subequations}
Here $x^\m$ are local coordinates on $X$, 
the one-forms $\{ E^\a \}$ constitute the {\it vielbein}, and $\O$ is called the {\it connection}.
Associated with  a group element $g \in G$ is  the transformation
\bea
x ~\to ~ x'=g\cdot x \quad \Longleftrightarrow \quad s(x) ~\to~ s(x') = g s(x) {\bm h}^{-1}(g,x)
\eea
which leads to: $ s^{-1} {\rm d} s ~ \to ~  {\bm h}\big( s^{-1} {\rm d} s \big)  {\bm h}^{-1} 
-{\rm d} {\bm h}\, {\bm h}^{-1}$, and hence 
\bea
E~ \to ~  {\bm h}E  {\bm h}^{-1} ~, \qquad 
\O  ~ \to ~  {\bm h}\O  {\bm h}^{-1} 
-{\rm d} {\bm h}\, {\bm h}^{-1}~.
\eea
The vielbein is seen to transform covariantly under $G$, while the transformation law 
of $\O$ includes an inhomogeneous piece typical of gauge fields.

Let $ \vf (x)$ be a field over $X$ with the group transformation law:
\bea
\vf (x) ~\to~\vf'(x') = {\bm h} (g,x)  \vf(x) ~, 
\eea
where, for simplicity of notation,  ${\bm h} (g,x) $ stands for   $R\Big({\bm h} (g,x) \Big)$.
The covariant derivative of $\vf$ is defined as follows:
\bea
{ \cD \vf :=  ({\rm d} +\O ) \vf = E^\a \, \cD_\a\vf }~, \qquad 
\cD_\a \vf := (E_\a +\O_\a ) \vf~.
\label{CovDer1}
\eea
Here $\{ E_\a = E_\a{}^\m (x) \pa_\m \}$ is the dual basis of $\{ E^\a ={\rm d} x^\m E_\m{}^\a (x) \}$, 
that is  
\bea
E_\a{}^\m (x) E_\m{}^\b (x) =\d_a{}^\b \quad \Longleftrightarrow \quad
E_\m{}^\a (x) E_\a{}^\n (x) =\d_\m{}^\n~.
\non
\eea

It should be remarked that the coset representative $s(x)$ is not uniquely defined.
The intrinsic freedom in its choice is described by gauge transformations
\bea
s(x) ~ \to ~ \widetilde{s}(x) := s(x) \, \ell^{-1} (x)~, \qquad \ell(x) \in H~, 
\eea
with $\ell (x)$ completely arbitrary. Under such a transformation, 
the geometric objects and fields change as follows:
\begin{subequations}
\bea
 {\bm h} (g,x)  ~& \to & ~ \widetilde{\bm h} (g,x) = \ell ( g \cdot x) {\bm h} (g,x)  \ell^{-1}(x)~,
 \\ 
E~  & \to &~  \widetilde{E}= {\ell} E  {\ell}^{-1} ~, \quad 
\O  ~ \to ~ \widetilde{\O} = {\ell}\O  {\ell}^{-1} 
-{\rm d} {\ell}\, {\ell}^{-1}~, \\
\vf ~ & \to &~  \widetilde{\vf}  = \ell  \vf~.
\eea
\end{subequations}

\subsection{Flat superspace geometry}
We can now apply the general formalism developed above to the case of 
the $\cN$-extended superspace ${\mathbb M}^{4|4\cN} $, using the following correspondence:
\bea
\phantom{|}\quad 
\begin{array}{|c|c|c| c| c| c| c| c|}
\hline
\phantom{\Big|}~X ~ {} & ~ G ~& ~ H ~& g &h &x^\m &s(x) &{\bm h} (g,x)\\
  \hline
\phantom{\Big|} {\mathbb M}^{4|4\cN} ~ &~    {\mathfrak P}(4|\cN)  ~  & ~ {\rm SL}(2,{\mathbb C}) ~&
~s(b,{\bm \ve} ) h(M) ~
& ~h(M) ~ & ~ z^M ~
&~s(z)  ~&~h(M) ~\\
\hline
\end{array}
\qquad {}
 \non
\eea
Here we have denoted 
$z^M = (x^m, \q^\m_\imath , {\bar \q}_{\dt \m}^\imath )$.
It is important to point out that $h(M$), which corresponds to ${\bm h} (g,x) $ 
in the case under consideration, has no explicit dependence on the superspace coordinates $z^M$.
It only remains to identify elements of the super-Poincar\'e algebra that correspond
to the generators $\cT_\a $ and $\cT_i $: 
\bea
\cT_\a ~ \to ~ \cT_A := ( P_a,   Q^i_\a , {\bar Q}^{\dt \a}_i  )~, 
\qquad \cT_i ~ \to ~ J_{ab} ~.
\non
\eea
As a result, we can read off the Maurer-Cartan form \cite{AV}
\bea
s^{-1} {\rm d} s  
&=& \left(
\begin{array}{c | c ||c}
0  ~& ~ 0~&~0   \\
\hline
-{\rm i}\,\tilde{\bm e}  ~& ~0~&~ 2{\rm d} {\bar \q} \\
\hline
\hline
2{\rm d} \q ~& ~0~&~0
\end{array}
\right)  ~, \qquad 
 e^a:= {\rm d} x^a + {\rm i} \big(\q_i\s^a {\rm d} {\bar \q}^i 
-{\rm d} \q_i \s^a {\bar \q}^i \big)~.~~~~
\eea
In particular, for the vielbein and connection we get
\bea
 e^A = {\rm d} z^M e_M{}^A( z) = \big( e^a ,\, {\rm d}\q^a_i , \, {\rm d}{\bar \q}_{\dt \a}^i \big)~,~~
 \quad  \O=0~.
\eea
The components of the vielbein, $e^A$, comprise the {supersymmetric one-forms}, 
i.e. those one-forms which are invariant under the supersymmetry transformations.

${}$Following (\ref{CovDer1}), 
for the covariant derivatives we obtain
\bea
\cD ={\rm d} \equiv {\rm d} z^M \frac{\pa}{\pa z_M } = e^A D_A ~, \qquad 
{ D_A = \big( \pa_a, D_\a^i, {\bar D}^{\dt \a}_i \big)}  ~,
\eea  
where the spinor covariant derivatives have the form:
\bea
D^i_\a = \frac{\pa}{\pa \q^{\a}_i}
+ {\rm i} \,(\s^b )_{\a \dt \b} \, {\bar \q}^{\dt \b i}\, \pa_b
  ~,  \qquad
{\bar D}_{\dt \a i} =
- \frac{\pa}{\pa {\bar \q}^{\dt \a i}} 
- {\rm i} \, \q^\b _i (\s^b )_{\b \dt \a} \,\pa_b ~. 
\label{SCD}
\eea
In the $\cN=1$ case, we denote $z^M = (x^m, \q^\m , {\bar \q}_{\dt \m} )$ and
$D_A = \big( \pa_a, D_\a, {\bar D}^{\dt \a} \big)$.

\subsection{Superfields}

\noindent
In accordance with (\ref{IndRep1}),  a tensor superfield $\cW(z)$, {\it with all indices suppressed},  
is defined to transform under the super-Poincar\'e group as follows:
\bea
g=s(b,{\bm \ve}) h(M): \quad \cW(z) \quad \longrightarrow \quad
\cW'(z') =  R(M) \cW(z)~, 
\label{IndRep2}
\eea
with $R$ a finite-dimensional representation of ${\rm SL}(2,{\mathbb C})$.
The important concept of superfields was introduced by Salam and Strathdee  \cite{SS}.

In the case of an infinitesimal supersymmetry transformation ($g=s(0,{\bm \ve})$), eq.
(\ref{IndRep2}) gives
\bea
 \d \cW := \cW'(z) -\cW(z) = 
 {\rm i} \,\big(  \e^\a_i \, { Q^i_\a} 
+ {\bar \e}^i_{\dt \a} \, {{\bar Q}^{\dt \a}_i } \big)\, \cW~,
\label{3.20}
\eea
where the {supersymmetry generators} have the form:
\bea
Q^i_\a = {\rm i} \,\frac{\pa}{\pa \q^{\a}_i}
+ (\s^b )_{\a \dt \b} \, {\bar \q}^{\dt \b i}\, \pa_b
  ~, 
\qquad
{\bar Q}_{\dt \a i} =
-{\rm i} \, \frac{\pa}{\pa {\bar \q}^{\dt \a i}} 
- \q^\b _i (\s^b )_{\b \dt \a} \,\pa_b
\label{SUSYg}
~. 
\eea
If $\cW(z)$ is a {tensor superfield}, then 
${D_A} \cW(z)$ 
is also a {tensor superfield}.
This implies that the covariant derivatives commute 
with  the supersymmetry transformations,
\bea
[ D_A ,  \e^\b_j \, { Q^j_\b} ]= [D_A , {\bar \e}^j_{\dt \b} \, {{\bar Q}^{\dt \b}_j }]=0~.
\eea

The spinor covariant derivatives obey the following anti-commutation relations:
\bea
\{D^i_{\a}  ,  D^j_{ \b} \} &=&
\{{\bar D}_{\dt \a i}  ,  {\bar D}_{\dt  \b j} \} = 0~,  \qquad
\{ D_\a^i  ,  {\bar D}_{\dt  \b j} \} =- 2 {\rm i} \d^i_j\,(\s^c)_{\a \dt \b} \,\pa_c~. 
\label{323}
\eea

\subsection{Chiral superfields}
Let us return to the coset representative (\ref{s(z)})
and consider its first  $(4+\cN)\times 2$ block-column 
\bea
{  \cC \big(x_{(+)}, \q\big)} =\left(
\begin{array}{c   }
\mathbbm{1}_2     \\
-{\rm i}\,\tilde{\bm x}_{(+)}    \\
2\q 
\end{array}
\right)  ~, \qquad  x^m_{(+)}:= x^m + {\rm i} \,\q_i \s^m {\bar \q}^i~. 
\eea
In accordance with (\ref{Gaction}),  
the  super-Poincar\'e transformation law of $\cC \big(x_{(+)}, \q\big)$ is:
\bea
\cC (x_{(+)}, \q) ~ \to ~ \cC (x'_{(+)}, \q') :=  g
 \,\cC (x_{(+)}, \q) \,M^{-1}~, \qquad g= s(b, {\bm \ve}) \,h (M)~.~~~
\eea
It follows that the variables $x^m_{(+)} $ and $\q^\a_i$ transform amongst themselves
(that is, they do not mix  with ${\bar \q}^{\ad i}$)  
under ${\mathfrak P}(4| \cN )$. 
This means that all superfields, which  depend on  $x^m_{(+)} $ and $\q^\a_i$ {only},
preserve this property under the super-Poincar\'e group: 
\bea
 \F (z):= \vf ( x_{(+)}, \q) \quad \Longrightarrow \quad 
\F'(z) =R(M) \F(g^{-1} \cdot z) =  \vf '( x_{(+)}, \q) ~.
\eea
Such superfields are singled out by the following first-order differential constraints
\bea
{\bar D}_{\dt \a i} \F =0 \quad \Longleftrightarrow \quad 
\F(x,\q , \bar \q) =  {\rm e}^{  {\rm i} \q_i\s^m {\bar \q}^i \pa_m} \, \vf(x,\q) 
\eea
and are called {\it chiral}.

The $\cN=1$ chiral scalar supermultiplet was discovered by Wess and Zumino \cite{WZ} 
in a component form, and some time later re-cast in superspace. 
Chiral superfields are indispensable in the context of $\cN=1, ~2$ supersymmetric theories.

\subsection{Supersymmetric action principle} 
In order to construct supersymmetric field theories, we have to learn how 
to generate supersymmetric invariants. For this, an indispensable mathematical 
concept is that of Berezin integral \cite{Berezin}. 

Consider a function $f(\q) $ of one Grassmann variable $\q$ or, equivalently, 
a function over ${\mathbb R}^{0|1}$. Integration over   ${\mathbb R}^{0|1}$ 
is defined by 
\be
\int {\rm d}\q \,f(\q) := \frac{\rm d}{\rm d \q} \,f(\q)
\equiv \frac{\rm d}{\rm d \q}\, f(\q) \Big|_{\q=0}~.
\ee
This definition can be immediately generalized to define integration over  ${\mathbb R}^{0|q}$.
Finally, in conjunction with  the standard notion of integration over ${\mathbb R}^p$, we can 
define integration over a superspace ${\mathbb R}^{p|q}$ as a multiple integral.  
A detailed discussion can be found, e.g., in \cite{BK}.

In the case of $\cN=1$ supersymmetry, the construction of the most general supersymmetric actions 
turns out to be almost trivial. 
Let $L(z)$ be a {real} {scalar superfield}. Then 
\bea
S := \int {\rm d}^4x {\rm d}^2\q {\rm d}^2 {\bar \q} \, L 
\label{N=1SUSYaction}
\eea
is invariant under the $\cN=1$ super-Poincar\'e group.
To prove the invariance of $S$, we note that it can be represented in the following equivalent forms:
\bea
S
&=& \frac{1}{16} \int {\rm d}^4x \, D^\a {\bar D}^2 D_\a L \Big|_{\q=0}
= \frac{1}{16} \int {\rm d}^4x \, {\bar D}_{\dt \a} D^2 {\bar D}^{\dt \a} L \Big|_{\q=0} ~,
\label{N=1SUSYaction2}
\eea
where we have made use of  the identity
\bea
D^\a {\bar D}^2 D_\a = {\bar D}_{\dt \a} D^2 {\bar D}^{\dt \a} ~, \quad
\qquad D^2 :=D^\a D_\a~, \quad {\bar D}^2 := {\bar D}_{\dt \a} {\bar D}^{\dt \a} ~.
\label{identityD4}
\eea
The proof goes as follows:
\bea
\d_{\rm SUSY} S &=& \phantom{-}\frac{\rm i}{16} \int {\rm d}^4x \, D^\a {\bar D}^2 D_\a 
\Big( \e Q + {\bar \e}{\bar Q} \Big)L \Big|_{\q=0} \non \\
&=&\phantom{-}
 \frac{\rm i}{16} \int {\rm d}^4x \, \Big( \e Q + {\bar \e}{\bar Q} \Big)
D^\a {\bar D}^2 D_\a 
L \Big|_{\q=0} \non \\
&=& -\frac{1}{16} \int {\rm d}^4x \, \Big( \e D + {\bar \e}{\bar D} \Big)
D^\a {\bar D}^2 D_\a 
L \Big|_{\q=0}
= \int {\rm d}^4x \,
 \pa_m f^m=0~,~~~~~
\label{3.32}
\eea
for some field $f^m(x)$.
Here we have made use of ({\it i}) the  explicit form of the spinor covariant 
derivatives (\ref{SCD}) and the supersymmetry generators (\ref{SUSYg}), 
as well as ({\it ii}) the anti-commutation relations (\ref{323}).

Along with the representations (\ref{N=1SUSYaction2}), the action (\ref{N=1SUSYaction})
can  also be written as
\bea
S
&=& \frac{1}{16} \int {\rm d}^4x \, D^2 {\bar D}^2 L \Big|_{\q=0}
= -\frac{1}{4} \int {\rm d}^4x \, D^2  L_{\rm c} \Big|_{\q=0}~, \quad 
L_{\rm c} :=  -\frac{1}{4}{\bar D}^2 L ~.~~~~~
\eea
The superfield $L_{\rm c}$ introduced can be seen to be chiral. This exercise 
leads to a new procedure to construct $\cN=1$ supersymmetric invariants. 
Given a chiral scalar $\cL_{\rm c}$, $ {\bar D}_{\dt \a} \cL_{\rm c}=0$, 
the functional
\bea
S_{\rm c} := \int {\rm d}^4x {\rm d}^2\q  \, \cL_{\rm c} 
\label{N=1SUSYaction3}
\eea
is invariant under the $\cN=1$ super-Poincar\'e group.

The above simple rules of constructing supersymmetric invariants
can be readily generalized to the case $\cN >1$.
 However, it turns out that this does not allow one to obtain 
the most interesting actions.

\section{Nonlinear sigma-models in $\cN = 1$ superspace}
\setcounter{equation}{0}
In four space-time dimensions, nonlinear sigma-models can possess two types
of supersymmetry: (i) $\cN=1$ supersymmetry or (ii) $\cN=2$ supersymmetry.\footnote{Only 
in these cases one can define a {\it scalar supermultiplet} comprising fields
of spin 0 and 1/2. The $\cN=2 $ scalar supermultiplet is also called a {\it hypermultiplet}.}
Here we review their formulations  in terms of $\cN=1$ chiral superfields.
In what follows, for the Grassmann integration measure 
we will use the notation $  {\rm d}^4 \q := {\rm d}^2 \q {\rm d}^2{\bar \q} $.

\subsection{$\cN = 1$ supersymmetric nonlinear sigma-models}
In 1979,  Zumino \cite{Zumino} put forward the following $\cN=1$ supersymmetric theory:
\bea
S&=& \int {\rm d}^4 x {\rm d}^4 \q  \,
 K\big(\F^a, {\bar \F}^{\bar{b}}\big)~, 
\qquad {\bar D}_{\dt \a} \F^a =0
\label{Zumino-sigma}
\eea
with the dynamical variables being $n$ {chiral scalar} superfields, $\F^a (z)$, 
and their complex conjugates,  ${\bar \F}^{\bar{b}} (z)$.
The above action is obtained from that describing $n $ free massless  scalar multiplets \cite{WZ} 
by replacing its quadratic Lagrangian, 
$ K_0\big(\F, {\bar \F} \big) = \d_{a \bar b} \, \F^a {\bar \F}^{\bar{b}}$,
with an arbitrary real analytic function. 
A key result of the work of \cite{Zumino} 
is  the geometric interpretation of the theory  (\ref{Zumino-sigma}) it provided.
It demonstrated that the {Lagrangian}
 $ K\big(\F^a, {\bar \F}^{\bar{b}}\big)$ can be interpreted as 
 the   {\it K\"ahler potential} of a {\it K\"ahler manifold} $ \cM$, 
 parametrized by local complex coordinates $\F^a$, 
with the following {\it K\"ahler metric}:
\bea 
g_{a \bar b} (\F, \bar \F ) &=& \frac{\pa^2 K}{\pa \F^a \pa {\bar \F}^{\bar b} } \equiv 
K_{a \bar b}~, \qquad g_{a  b}=g_{\bar a \bar b} =0~.
\label{KahlerMetric}
\eea
Here and in what follows, we  use the notation:
\bea
K_{a_1 \dots a_p \,{\bar b}_1 \dots {\bar b}_q } &:=& 
\frac{\pa^{p+q} K}{\pa \F^{a_1} \dots \pa \F^{a_p}\, {\bar \F}^{{\bar b}_1} \dots {\bar \F}^{{\bar b}_q}}~.
\eea

As is well-known, the metric on $\cM$ can locally be expressed in terms of a single function, 
eq. (\ref{KahlerMetric}), due to the fact that the K\"ahler form 
\bea
\o = {\rm i} \, g_{a \bar b} \,{\rm d}  \F^a \wedge  {\rm d}{\bar \F }^{\bar b}
\eea
is closed, ${\rm d} \o=0$. The metric (\ref{KahlerMetric}) does not change under
a {\it K\"ahler transformation} 
\bea
 K\big(\F, {\bar \F} \big) \quad \longrightarrow \quad 
  K\big(\F, {\bar \F} \big) + \L(\F) +{\bar \L} (\bar \F )~, 
\label{KahlerInvariance}
\eea
with $\L (\F )$ an arbitrary holomorphic function. For the above interpretation of 
 the theory  (\ref{Zumino-sigma})  to be correct, 
the  action functional must be invariant under arbitrary K\"ahler transformations.
This indeed follows from the facts that (i) the action can be represented, due to   (\ref{identityD4}), as
\bea
 S = \int {\rm d}^4x \,{\cL} ~,\qquad 
 16\,{\cL} :&=& {D^\a {\bar D}^2 D_\a }K \big|_{\q=0} 
= {\bar D}_{\dt \a} D^2 {\bar D}^{\dt \a} K \big|_{\q=0} ~;
\label{4.6}
\eea
and (ii) the space of chiral superfields has a ring structure, that is  
\be
{\bar D}_{\dt \a} \F^a =0 \quad \longrightarrow \quad {\bar D}_{\dt \a} \L (\F ) =0~.
\ee

The above analysis actually shows that the component Lagrangian, $\cL$, in 
(\ref{4.6}) is invariant under arbitrary K\"ahler transformations (\ref{KahlerInvariance}).
This property allows us to demonstrate that the theory is independent of a choice 
of local coordinates in the target space. Specifically, 
if $\{U_{(i)} , \F_{(i)} \}$ is an atlas on $\cM$, and $K_{(i)}( \F_{(i)}, \bar  \F_{(i)} )$ is  the local 
K\"ahler potential corresponding to 
the chart $U_{(i)}$, then one and the same point  $p \in \cM$ can belong to several charts.
In the intersections of two charts, $U_{(i)}$ and $U_{(j)}$, we have 
\bea
K_{(j)}( \F_{(j)}, \bar  \F_{(j)} )=K_{(i)}( \F_{(i)}, \bar  \F_{(i)} )
+ \Big[ \L (\F_{(i)} ) +  {\rm c.c.} \Big] ~, \quad 
\F^a_{(j)} = f^a ( \F_{(i)} )~, ~~~
\eea
for some holomorphic functions $\L (\F) $ and $f^a (\F)$. From here we can see 
that the Lagrangian $ \cL$ is indeed independent of 
the choice of $K_{(i)}$ made. 

Let us turn to computing the component Lagrangian.
Introduce the {component fields } of $\F^a (z)$ by the rule:
\bea
\F^a(x,\q , \bar \q) =  {\rm e}^{ {\rm i} \q\s^m {\bar \q} \pa_m} \, \Big\{
{  \vf^a (x)} 
+\q\, { \j^a (x)} + \q^2 { F^a(x)} \Big\}~.
\eea
Here $\vf^a $ and $F^a$ are {complex scalar} fields, while $\j^a_\a $ a 
{spinor} field.
Direct calculations lead to 
\bea
{ \cL} &=& 
- g_{a \bar b}(\vf ,\bar \vf)  \Big( \pa^m \vf^a \,
\pa_m {\bar \vf}^{\bar b} 
+\frac{\rm i}{4} \j^a  \s^m  \!\stackrel{\leftrightarrow}{\nabla}_m {\bar \j}^{\bar b}\Big)
+g_{a \bar b}(\vf ,\bar \vf) \,\cF^a {\bar \cF}^{\bar b} \non \\
&&
+ \frac{1}{16} R_{a\bar b c \bar d}(\vf,\bar \vf) \, \j^a \j^c \,{\bar \j}^{\bar b} {\bar \j}^{\bar d}~,
\eea 
where $\nabla_m \j^a$ denotes the covariant derivative of $\j^a$, 
\bea
\nabla_m \j^a := \pa_m \j^a + (\pa_m \vf^b ) \,\G^a_{bc} (\vf, \bar \vf) \,\j^c ~,
\eea
and we also define
\bea
\cF^a := F^a -\frac{1}{4} \G^a_{bc} (\vf, \bar \vf) \, \j^b \j^c~.
\eea
${}$Finally, $\G^a_{bc}(\vf, \bar \vf) $ and $R_{a\bar b c \bar d}(\vf,\bar \vf) $ denote the Christoffel symbols 
and the Riemann tensor associated with the K\"ahler metric $g_{a\bar b}(\vf, \bar \vf )$,
\bea
\G^a_{bc} = g^{a\bar d} K_{bc \bar d}~, \qquad 
R_{a\bar b c \bar d} = K_{ac \bar b \bar d} - g_{e \bar f} \, \G^e_{ac} \, \G^{\bar f}_{\bar b \bar d}~.
\eea

The equations of motion for the ${\bar F}$s are: 
\bea
\cF^a =0 \quad \longleftrightarrow \quad F^a =\frac{1}{4} \G^a_{bc} (\vf, \bar \vf) \, \j^b \j^c~.
\eea
The fields $F^a $ and their conjugates ${\bar F}^{\bar a}$ appear in the action without derivatives.
When their equations of motion hold, they become functions of other fields. 
Their sole role is to have supersymmetry {linearly realized}. 
Such fields are called {\it auxiliary}.

\subsection{$\cN = 2$ supersymmetric nonlinear sigma-models}
\label{N=2-->N=1}
How to construct $\cN=2$ supersymmetric nonlinear sigma-models? 
A possible approach is to work in terms of $\cN=1$ superfields. 
In such a setting,  one starts  from the general $\cN=1$ supersymmetric nonlinear 
sigma-model \cite{Zumino}
\bea
S&=& \int {\rm d}^4 x {\rm d}^4 \q  \, {\mathfrak K}\big(\F^a, {\bar \F}^{\overline{b}}\big)~, 
\qquad {\bar D}_{\dt \a} \F^a =0~,
\label{N=2SMN=1}
\eea
which is associated with some K\"ahler manifold $\cM$, 
and look for those  target space geometries which are compatible with  an additional 
hidden   supersymmetry.

As a first step, it is necessary to make an educated guess regarding the explicit  form of 
a second supersymmetry.  
A correct ansatz was proposed in \cite{LR,HKLR}.  
It is defined modulo an irrelevant  {\it trivial symmetry} transformation
(that is proportional to the equations of motion), 
such as
\be
\d \vf^i = \G^{ij} \, \frac{\d S[\vf] }{\d \vf^j}~, \qquad \G^{ij}=-\G^{ji} 
\ee
that any theory $S[\vf]$ of bosonic fields $\vf^i$ possesses.
The   second supersymmetry is:
\bea
\d \F^a = \hf {\bar D}^2 \Big( {\bar \e} (\bar \q) \, {{\bar \O}^a} \Big)~, \qquad 
\d {\bar \F}^{\bar a}  = \hf D^2 \Big( { \e}(\q) \,
{\O^{\bar a} }\Big) ~,
\label{4.16}
\eea 
for some functions
$\O^{\bar a} =\O^{\bar a} \big(\F, \bar \F\big)$
associated with  the K\"ahler manifold $\cM$. Here the  transformation parameter 
$\e (\q)$ has the form:  
\bea
 \e (\q) = \t + \e^\a \q_\a ~, \qquad \t , \e^\a ={\rm const}~, 
\label{4.17}
\eea
where $\e^\a$ is the supersymmetry parameter,
while $\t $  generates a central charge transformation. 
Actually, the latter transformation should be a trivial symmetry, for it is not present in the supersymmetry 
algebra (\ref{2.23c}) and (\ref{2.23d}).  However, it is natural to keep it in (\ref{4.17}), 
because such a transformation is generated, off the mass shell,  by commuting 
the first and second supersymmetries. 

There are two simple observations to justify the fact  that the ansatz (\ref{4.16}) is indeed general. 
Firstly, since $\d \F^a$ must be chiral, it can be represented $\d \F^a = {\bar D}^2 (\dots )$.
Secondly, we can assign dimension zero to $\F^a$, and then any function ${\bar \O}^{ a} \big(\F, \bar \F\big)$
is also dimensionless. 
The mass dimensions of $\e^\a$, $\q^\a$ and ${\bar D}^2$ are, respectively,  $-1/2$, $-1/2$ and $+1$.
These observations lead to (\ref{4.16}). 
To be more precise, it is possible to  deform the variation $\d \F^a $ given in (\ref{4.16}) 
by adding a term proportional to  $ { \e}^{ \a} {\q}_{\a}{\bar D}^2 {\bar \O}^a$.
However, the latter proves to generate a trivial symmetry (see, e.g., \cite{K09} for more details),
and therefore can be ignored.
In section 9, we show that the transformation law  (\ref{4.16}) naturally follows from an
off-shell formulation for $\cN=2$ supersymmetric nonlinear sigma-models.

The next steps should be to analyze the implications of the 
requirements that (i) the action (\ref{N=2SMN=1}) be invariant under the transformations
 (\ref{4.16}); and (ii) the first and second supersymmetry transformations form 
 the $\cN=2$ super-Poincar\'e algebra on the mass-shell.
This analysis was carried out in \cite{HKLR}, and here we only summarize the  results
obtained.
\begin{itemize}
\item
The action (\ref{N=2SMN=1}) is invariant under the transformations  (\ref{4.16})  if 
the following conditions hold:
\bea
{ \o}_{ b  {c} } :=  g_{b \overline{a} }\, { \O}^{\bar a}{}_{,{c} }
 =  -  { \o}_{ {c}  {b} }~,
\label{hol-two-form}
\eea
and 
\begin{subequations}
\bea
{ \o}_{ {b}  {c} \,, \bar a } :=  \pa_{\bar a} { \o}_{ {b}  {c}  }
=\nabla_{\bar a}  { \o}_{ {b}  {c}  }&=&0 ~, \label{4.19a}\\
\nabla_{ a}  { \o}_{ {b}  {c}  }&=&0 ~.
\label{4.19b}
\eea
\end{subequations}
It can be shown that $\o_{bc} (\F) $ is a globally defined {\it holomorphic two-form} on $\cM$.\footnote{This
follows from the fact that, {\it on the mass shell}, the variations $\d \F^a$ and $\d {\bar \F}^{\bar a} $ in  (\ref{4.16}) should 
constitute a vector field on $\cM$.}
Eqs. (\ref{4.19a}) and (\ref{4.19b})
mean that the two-form  $\o_{bc} $ is {\it covariantly constant}, 
and therefore the target space $\cM$ is a manifold of restricted holonomy.

\item
The first and the second supersymmetries form the $\cN=2$ super-Poincar\'e 
algebra (with $ i,j=\1, \2$, where the values of isospinor indices are underlined for later convenience),
\bea 
\{Q^i_{\a} \, , \, Q^j_{ \b} \} =
\{{\bar Q}_{\dot \a i} \, , \, {\bar Q}_{\dot  \b j} \} = 0~,   \qquad 
\{ Q_\a^i \, , \, {\bar Q}_{\dot  \b j} \} =
2 \d^i_j\,(\s_c)_{\a \bd} \,P^c~, ~~~
\eea
{\it on the equations of motion} if
\bea
 {\bar \O}^a{}_{, \bar c}  \, { \O}^{\bar c}{}_{,  b}
=- \d^a{}_b~.
\eea
\end{itemize}
A detailed derivation of the above results can be found in \cite{K09}.

Denote by $ J_3 $ the complex structure chosen on the target space $\cM$,
\bea
J_3 = \left(
\begin{array}{cc}
{\rm i} \, \d^a{}_b  ~ & ~ 0 \\
0 ~ &   -{\rm i} \, \d^{\bar a}{}_{\bar b}  
\end{array}
\right)~.
\label{ComplexSturcture3} 
\eea
It follows from the previous results that there exist
two more {covariantly constant complex structures} 
\bea
J_1 &:=& \left(
\begin{array}{cc}
0  ~ & ~ {g}^{a \bar c} {\bar \o}_{\bar c \bar b} \\
{g}^{ \bar a c } { \o}_{ c  b}
 ~ &   0
\end{array}
\right)~,
\qquad 
J_2 := \left(
\begin{array}{cc}
0  ~ &  {\rm i}\,   {g}^{a \bar c} {\bar \o}_{\bar c \bar b}  \\
-{\rm i}\,   {g}^{ \bar a c } { \o}_{ c  b}~ &   0
\end{array}
\right)
\label{ComplexSturcture1-2}
\eea
such that {(i)} $\cM$ is K\"ahler with respect to each of them; and  
{(ii)}
the operators $J_A = (J_1,J_2,J_3) $ form the quaternionic algebra:
\bea
J_A \,J_B = -\d_{AB} \, {\mathbbm 1} + \ve_{ABC}J_C~.
\eea
Therefore the target space $\cM$ is a hyperk\"ahler 
manifold.

Given a hyperk\"ahler space $(\cM , g, J_A)$, 
we pick one of its complex structures, say $J_3$, and introduce complex 
coordinates $\f^a$ compatible with it. In these coordinates, $J_3$ has the form 
(\ref{ComplexSturcture3}).  Then, two other complex structures, $J_1$ and $J_2$, 
are given by eq. (\ref{ComplexSturcture1-2}).
The matrix elements of  $J_1$ and $J_2$ are determined by the holomorphic two-form, 
eq. (\ref{hol-two-form}), from which we cannot directly read off 
the functions ${\bar \O}^a$ and $\O^{\bar a}$ appearing in (\ref{4.16}), but only their 
partial derivatives.
Ref. \cite{HKLR} presented the following explicit expression for ${\bar \O}^a$:
\bea 
{\bar \O}^a = \o^{ab}  \big(\F \big) {\mathfrak K}_b \big(\F , \bar \F \big)~, \qquad
{\mathfrak K}_b \big(\F , \bar \F \big):= \frac{\pa}{\pa \F^b} {\mathfrak K}\big(\F , \bar \F \big)~.
\eea
Although ${\bar \O}^a $ changes under the K\"ahler transformations as  
\bea
{\mathfrak K} \big(\F , \bar \F \big)  ~& \to& ~
{\mathfrak K}\big(\F , \bar \F \big) +  \L \big(\F  \big) + {\bar \L} \big( \bar \F \big)   ~, \non \\ 
 \o^{ab} (\F) {\mathfrak K}_b \big(\F , \bar \F \big) ~& \to & ~ 
 \o^{ab}(\F)   {\mathfrak K}_b \big(\F , \bar \F \big) +  \o^{ab}(\F)   \L_b (\F)~,
\eea 
the supersymmetry transformation $ \d \F^a = \hf {\bar D}^2 \big( {\bar \e} \,{\bar \O}^a \big)$ remains invariant.

The Lagrangian of the $\cN=2$ supersymmetric sigma-model, eq. (\ref{N=2SMN=1}), 
is the {hyperk\"ahler potential} of $\cM$.

\section{$\cN = 2$ superspace with auxiliary  dimensions}
\setcounter{equation}{0}
As with the component  ($\cN=0$) formulation 
for general $\cN=2$ supersymmetric nonlinear sigma-models \cite{A-GF,BW}, 
their formulation in terms of $\cN=1$ superfields described above
is just an {\it existence theorem}.  The $\cN=1$ formulation has two major drawbacks:

\begin{itemize}
\item It is not suitable from the point of view  
of generating $\cN=2$ supersymmetric nonlinear sigma-models;

\item It  provides {\it no} insight  from the point of view  
of constructing $\cN=2$ superconformal  nonlinear sigma-models.
\end{itemize}
To overcome the drawbacks of the $\cN=1$ formalism
is hardly possible without making use of $\cN=2$ superspace techniques.
However, in the early 1980s there emerged 
a conceptual problem concerning such techniques. 
It was realized that standard multiplets defined in the {conventional $\cN=2$ superspace 
${\mathbb M}^{4|8}$} are not suitable (say, {\it too long}) for sigma-model 
constructions. A way out was to look for an extension of the conventional superspace.

The correct superspace setting was found in 1983--1984 independently by three groups 
who pursued somewhat different goals \cite{Rosly,GIKOS,KLR}. It is 
\bea
{\mathbb M}^{4|8}\times {\mathbb C}P^1={\mathbb M}^{4|8}\times S^2~.
\eea
Below, we will briefly discuss each of the three approaches mentioned.

\subsection{Isotwistor superspace}
\label{Iso-superspace}
In order to introduce the construction given in \cite{Rosly}, we should start from 
the algebra of $\cN=2$ spinor covariant derivatives ($i, j =\1, \2$): 
\bea
\{D^i_{\a} \, , \, D^j_{ \b} \} = 0~,
\quad 
\{{\bar D}_{\dt \a }^i \, , \, {\bar D}_{\dt  \b }^j \} = 0~,  \quad 
\{D^i_{\a} \, , \, \bar D_{ \dt \b }^j \} = 2{\rm i} \, \ve^{i j}\,
(\s^m )_{\a \dt \b} \,\pa_m ~.
\label{5.2}
\eea

${}$Following the  work\footnote{Rosly's approach \cite{Rosly} was inspired by earlier ideas 
due to Witten \cite{Witten}.} of
\cite{Rosly}, introduce an {\it isotwistor} 
$ v^i \in {\mathbb C}^2 \setminus \{0\}$ and define\footnote{See Appendix C for our convention 
to raise and lower isotwistor indices.} 
\bea 
{\mathfrak D}_{ \a} :=v_i \,{ D}^i_{ \a} ~, \qquad 
{\bar {\mathfrak D}}_{\dt \a} := v_i \,{\bar { D}}^i_{\dt \a}~, \qquad
v_i:= \ve_{ij}\,v^j~.
\label{5.3}
\eea
Then, the anti-commutation  relations (\ref{5.2}) imply that 
\bea
\{ {\mathfrak D}_{ \a} , {\mathfrak D}_{ \b} \} = \{{\mathfrak D}_{ \a} , {\bar {\mathfrak D}}_{\dt \b }\} 
=\{ {\bar {\mathfrak D}}_{\dt \a} , {\bar {\mathfrak D}}_{\dt \b} \}=0~.
\eea
These relations allow us to introduce a new type of 
superfields obeying the (Grassmann) {\it analyticity constraints}: 
\bea
 {\mathfrak D}_{ \a} \f = {\bar {\mathfrak D}}_{\dt \a } \f =0~, 
\qquad \f= { \f(z,v ,\bar v)}~, \qquad {\bar v}_i :=  (v^i )^*~.
\label{AnaCo}
\eea
Such a superfield depends of  half of the Grassmann coordinates.

It should be pointed out that the operators ${\mathfrak D}_{ \a} $ and ${\bar {\mathfrak D}}_{\dt \a} $, 
eq. (\ref{5.3}), 
are not complex conjugate of each other. However, they turn out to be conjugate 
with respect to the generalized conjugation define in subsection \ref{smile}.

The constraints
$ {\mathfrak D}_{ \a} \f = {\bar {\mathfrak D}}_{\dt \a } \f =0$
do not change if we replace ${ v^i \to c\, v^i}$, with $c\in {\mathbb C}^*$, 
in the definition of ${\mathfrak D}_{ \a} $ and ${\bar {\mathfrak D}}_{\dt \a}$.
It is natural to restrict our attention to those superfields
which (i) obey  the  constraints $ {\mathfrak D}_{ \a} \f = {\bar {\mathfrak D}}_{\dt \a } \f =0$ 
and (ii) only scale  under arbitrary re-scalings  of $v$: 
\bea
\f(z,c\, v, {\bar c}\, \overline{v})= c^{n_+} \, {\bar c}^{n_-} \,\f(z,v, \overline{v})~,
\qquad c\in {\mathbb C}^*
\eea
for some parameters $n_\pm$ such that ${ n_+ - n_-}$ is an integer. 
By redefining $ \f(z,v, \bar v) \to \f(z, v, \bar v)/  (v^\dagger v  ) ^{n_-}$,
we can always choose ${ n_-=0}$. 
Any superfield with the homogeneity property 
\bea
{\ \f^{(n)}} (z,c\, v, {\bar c}\, \overline{v})= c^{n}  \,\f^{(n)}(z,v, \overline{v})~,
\qquad c\in {\mathbb C}^*
\label{HomCon}
\eea
is said to have {weight $n$}.
A weight-$n$ {\it isotwistor superfield} is defined to obey the following properties:
\bea
 {\mathfrak D}_{ \a} \f^{(n)} &=& {\bar {\mathfrak D}}_{\dt \a } \f^{(n)} =0~, \quad
  \f^{(n)} (z,c\, v, {\bar c}\, \overline{v})= c^{n}  \,\f^{(n)}(z,v, \overline{v})~,
\quad c\in {\mathbb C}^* ~.
\label{5.8}
 \eea

We see that the isotwistor
$ v^i \in {\mathbb C}^2 \setminus\{0\}$ is   defined modulo the equivalence relation
$ v^i \sim c\,v^i$,  with $c\in {\mathbb C}^*$, {hence it parametrizes ${\mathbb C}P^1$}.
The isotwistor superfields introduced live in the space 
${\mathbb M}^{4|8}\times {\mathbb C}P^1$ 
which was called {\it isotwistor superspace} by Rosly and Schwarz \cite{RS}.

Given an isotwistor superfield $\f^{(n)}(v^i, {\bar v}_j)$, its complex conjugate 
\bea
{\bar \f}^{(n)} ({\bar v}_i, {v}^j) 
:=
\overline{ \f^{(n)}(v^i, {\bar v}_j) }
\eea
is no longer isotwistor, for it  satisfies neither  the constraints (\ref{AnaCo}) nor the homogeneity 
condition (\ref{HomCon}). This is completely similar to the situation with the chiral superfields.
However, there is a fundamental difference between the isotwistor and chiral superfields: 
for the former one can define a modified conjugation that maps any isotwistor superfield
$\f^{(n)}(v^i, {\bar v}_j)$ into an isotwistor one  $\breve{\f}^{(n)}(v^i, {\bar v}_j)$ defined 
as a composition of the complex conjugation with the antipodal mapping\footnote{The smile conjugation 
is similar to the Dirac conjugation of four-component spinors defined as follows:  
$\J \to \overline{\J} := \J^\dagger \g^0$.}  
on $S^2$:
\bea
 \f^{(n)}(v^i, {\bar v}_j) \longrightarrow  {\bar \f}^{(n)} ({\bar v}_i, {v}^j) 
  \longrightarrow  {\bar \f}^{(n)} \Big({\bar v}_i \to -v_i, {v}^j \to {\bar v}^j\Big) =:\breve{\f}^{(n)}(v^i, {\bar v}_j)~.
~~~
\label{smile-iso}
\eea
The weight-$n$ isotwistor superfield $\breve{\f}^{(n)}(v^i, {\bar v}_j)$ is said to be the 
smile-conjugate of  $\f^{(n)}(v^i, {\bar v}_j)$. One can check that 
\bea
\breve{ \breve{\f}}^{(n)}(v, {\bar v}) =(-1)^n {\f}^{(n)}(v, {\bar v})~.
\eea
Therefore, if the weight $n$ is even, real isotwistor superfields can be defined, 
 $\breve{\f}^{(2m)}(v , {\bar v}) = {\f}^{(2m)}(v , {\bar v})$.

\subsection{Harmonic superspace approach}
We turn to a very brief discussion of the harmonic superspace approach pioneered by 
Galperin, Ivanov, Kalitsyn, Ogievetsky and Sokatchev \cite{GIKOS}.
A detailed account can be found, e.g., in the monograph \cite{GIOS}.

One can use the equivalence relation
$v^i \sim c\,v^i$,  with $c\in {\mathbb C}^*$, to switch to a   description in terms of  
the following { normalized}  isotwistors:
\bea
u^{+i} := \frac{v^i}{\sqrt{v^\dagger v}}~, \qquad u^-_i :=  \frac{{\bar v}_i}{\sqrt{v^\dagger v}}
=\overline{u^{+i}} 
\quad \Longrightarrow \quad 
{ \Big(u_i{}^- , u_i{}^+ \Big) \in {\rm SU(2)}}~.
\eea
The $u^\pm_i$ are called {\it harmonics}. They  are defined modulo the equivalence relation 
 $ u^\pm_i \sim \exp (\pm {\rm i}\a)\,u^\pm_i$,  with $\a \in {\mathbb R}$.
It is clear that  the harmonics parametrize the coset space
$ {\rm SU(2)/U(1)} \cong S^2$.

Associated with an isotwistor superfield  $\f^{(n)}(z,v, \overline{v})$
is the following superfield 
 \bea
 \vf^{(n)} (z,u^+,u^-) := \f^{(n)}\left(z, \frac{v}{ \sqrt{v^\dagger v} }, \frac{\bar v}{ \sqrt{ v^\dagger v} }\right)  =
 \frac{1}{ (\sqrt{v^\dagger v})^n } \f^{(n)}(z,v, \overline{v})
 \eea
obeying the  homogeneity condition
\bea
\vf^{(n)}(z, {\rm e}^{ {\rm i}\a}\, u^+, {\rm e}^{ -{\rm i}\a}\,u^-)
= {\rm e}^{ {\rm i}n\a} \,\vf^{(n)}(z,u^+, u^-)~.
\eea
The $\vf^{(n)}(z, u^\pm) $ is said to  have U(1) charge $n$. 

{Within the harmonic superspace approach, 
$\vf^{(n)}(z, u^\pm) $ is required to be a 
{\it smooth} charge-$n$ function over SU(2) or, equivalently, a {\it smooth} tensor field over 
the two-sphere ${ S}^2 $}. Such a superfield is called {\it analytic}.
It can be represented, say for $n\geq 0$, by a convergent Fourier series (see, e.g, \cite{Zhel}) 
\bea
\vf^{(n)}(z, u^\pm) = \sum_{p=0}^{\infty} 
\vf^{ (i_1 \dots i_{n+p}  j_1 \dots j_p )} (z)\,
u^+_{i_1} \dots u^+_{i_{n+p}} u^-_{j_1} \dots  u^-_{ j_p }  ~,
\eea
in which the coefficients $\vf^{ i_1 \dots i_{n+2p}   } (z)= \vf^{( i_1 \dots i_{n+2p}   )} (z)$
are ordinary $\cN=2$ superfields obeying first-order differential constraints that follow 
from (\ref{5.8}).
The beauty of this approach is that  the power of harmonic analysis can be used.

To construct supersymmetric theories,   a {\it supersymmetric action principle} is required.
In harmonic superspace, it 
includes integration over $S^2$ 
in addition to that  over the space-time and (half of) Grassmann variables.
Let $L^{(4)} (z,u^\pm ) $ be a real analytic superfield of U(1) charge $+4$, 
and 
\bea
\cL^{(4)} (z,v, \bar v) :=  (v^\dagger v)^2 \, L^{(4)} (z,u^+, u^-)
\eea
the corresponding weight-$n$ isotwistor superfield. Associated with $\cL^{(4)}$
is the following $\cN=2$ supersymmetric invariant:
\bea
S:= \int {\rm d}^4x \int {\rm d}^2\m \, \D^{(-4)} \cL^{(4)} (z,v, \bar v) \Big|_{\q={\bar \q} =0}~. 
\label{Harmonic action}
\eea
Here 
\bea
{\rm d}^2\m := \frac{\rm i}{2\p}\frac{ v_i {\rm d} v^i \wedge {\bar v}^j {\rm d} {\bar v}_j }{ (v^\dagger v)^2} 
= \frac{\rm i}{2\p} \frac{ v_i {\rm d} v^i \wedge {\bar v}^j {\rm d} {\bar v}_j }{ ({\bar v}_k v^k)^2} 
\eea
can be recognized as the usual measure on $S^2$. Indeed, introducing a complex 
(inhomogeneous) coordinate $\z$ in the north chart of ${\mathbb C}P^1$ as 
\bea 
 v^i = v^{\1} \,(1, \z) ~,\qquad \z:=\frac{v^{\2}}{v^{\1}} ~,\qquad\quad 
{ i=\1 ,\2}
\label{Zeta}
\eea
one obtains
\bea
{\rm d}^2\m =\frac{\rm i}{2\p} \frac{  {\rm d} \z  \wedge {\rm d} {\bar \z}  } {(1 +\z \bar \z)^2}~. 
\eea
The operator $\D^{(-4)} $ in (\ref{Harmonic action}) is 
\bea
\D^{(-4)} := \frac{1}{16} \nabla^\a \nabla_\a {\bar \nabla}_{\dt \b}  {\bar \nabla}^{\dt \b} ~, \qquad
\nabla_\a := \frac{1}{v^\dagger v} {\bar v}_i D^i_\a ~, \quad 
{\bar \nabla}_{\dt \b} := \frac{1}{v^\dagger v} {\bar v}_i {\bar D}^i_{\dt \b} ~.~~~
\eea

\subsection{Projective superspace approach}
\label{PSA}
The formation of the projective superspace  approach 
\cite{KLR,GHR,LR-projective1,LR-projective2,G-RRWLvU} 
has taken several years, from 1984 to 1990, although its key elements  already appeared 
in the work by  Karlhede,  Lindstr\"om and Ro\v cek \cite{KLR} on self-interacting $\cN=2$ 
tensor multiplets. The name `projective superspace' was coined in 1990 \cite{LR-projective2}.
Modern projective-superspace terminology appeared in the 1998 work \cite{G-RRWLvU} mostly
devoted to quantum aspects, along with important formal developments, of the approach.

In this approach, off-shell supermultiplets are described in terms of weight-$n$ isotwistor
superfields $Q^{(n)}(z,v)$,
\bea
 {\mathfrak D}_{ \a} Q^{(n)} &=& {\bar {\mathfrak D}}_{\dt \a } Q^{(n)} =0~, \quad
  Q^{(n)} (z,c\, v )= c^{n}  \, Q^{(n)}(z,v)~,
\quad c\in {\mathbb C}^* 
\label{5.21}
 \eea
which are { {\it holomorphic } over an {\it open domain} of
$ {\mathbb C}P^1$},
\bea
\frac{\pa}{\pa {\bar v}_i} \, Q^{(n)} =0~.
\eea
Such a superfield is called {\it weight-$n$ projective superfield}.\footnote{The terminology 
`weight-$n$ projective superfield' appears to be more appropriate than `degree-$n$ projective superfield' 
because in the superconformal case the parameter $n$ coincides with the superconformal weight of 
 $Q^{(n)}(z,v)$ \cite{K09}.}
There is no need to require  $Q^{(n)}(z,v)$
to be holomorphic over  ${\mathbb C}P^1$, for such a requirement  is not essential 
for the construction of projective-superspace actions. 

The $\cN=2$ {\it supersymmetric action principle} is formulated in terms of 
a Lagrangian $\cL^{(2)}(z,v)$ which is  a  real weight-2 projective superfield.
The action functional  {includes a closed contour integral in ${\mathbb C}P^1$}, 
along with  integration 
over Minkowski space and half of the Grassmann variables:
\bea
S:= - \frac{1}{2\p} \oint_{\g}  { v_i {\rm d} v^i }
\int {\rm d}^4x \, \D^{(-4)} \cL^{(2)} (z,v) \Big|_{\q={\bar \q} =0}~. 
\label{PAP}
\eea
Here $\g$ denotes a closed contour  in  ${\mathbb C}P^1$, $v^i(t)$,
parametrized by an evolution parameter $t$.
The action makes use of the following fourth-order differential operator:
\bea
\D^{(-4)} := \frac{1}{16} \nabla^\a \nabla_\a {\bar \nabla}_{\dt \b}  {\bar \nabla}^{\dt \b} ~, \quad
\nabla_\a := \frac{1 }{ (v,u)} 
{ u_i} D^i_\a 
~, \quad 
{\bar \nabla}_{\dt \b} := \frac{1 }{(v,u)} 
u_i{\bar D}^i_{\dt \b} ~,~~
\eea
where $(v,u):= v^i u_i$.
Here $u_i$ is a {fixed} isotwistor chosen to be arbitrary modulo 
the condition $(v,u) \neq 0$ along the integration contour.

Making use of the analyticity constraints obeyed by $\cL^{(2)} (z,v)$, one can show that the action 
is invariant under the $\cN=2$ super-Poincar\'e group. The proof is analogous to that considered earlier 
in the  $\cN=1$ case, eq. (\ref{3.32}).
The supersymmetry transformation acts on $\cL^{(2)}$ as follows:
\bea
 \d_{\rm SUSY} \cL^{(2)} =
 {\rm i} \,\big(  \e^\a_i \, { Q^i_\a} 
+ {\bar \e}^i_{\dt \a} \, {{\bar Q}^{\dt \a}_i } \big)\cL^{(2)} 
= {\rm i} \,\big(  \e_i  { Q^i} 
+ {\bar \e}^i {{\bar Q}_i } \big)\cL^{(2)} 
~,
\eea
compare with eq. (\ref{3.20}). Since the supersymmetry generators anti-commute with 
the spinor covariant derivatives, the variation of the actions is:
\bea
\d_{\rm SUSY} S&=& - \frac{\rm i}{2\p} \oint_{\g}  { v_i {\rm d} v^i }
\int {\rm d}^4x \,  \big(  \e_i  { Q^i} + {\bar \e}^i {{\bar Q}_i } \big)
\D^{(-4)} \cL^{(2)} (z,v) \Big|_{\q={\bar \q} =0} \non \\
&=& \frac{1}{2\p} \oint_{\g}  { v_i {\rm d} v^i }
\int {\rm d}^4x \,  \big(  \e_i  { D^i} + {\bar \e}^i {{\bar D}_i } \big)
\D^{(-4)} \cL^{(2)} (z,v) \Big|_{\q={\bar \q} =0} ~,~~~~
\label{5.26}
\eea
where we have made use of the explicit form of the supersymmetry generators 
and spinor covariant derivatives. 
Using the completeness relation 
\bea
\d^i_j = \frac{ v^iu_j -v_j u^i}{(v,u)}~,
\eea
the first term on the right can be transformed 
as follows:
\bea
 \e_i   D^i =
 v^i\e_i \nabla -   \frac{1}{(v,u)}\,
 u^i\e_i  
 {\mathfrak D} ~.
\label{5.29-new}
\eea
Here the first term does not contribute to (\ref{5.26}), since 
$\nabla_\a \D^{(-4)} ={\bar \nabla}_{\dt \a} \D^{(-4)} =0$. As to the second term, 
the operator $ {\mathfrak D}_\a$ can be pushed through $\D^{(-4)}$ in (\ref{5.26}) until it hits
$\cL^{(2)}$, which gives zero, due to (\ref{5.21}). In the process of pulling $ {\mathfrak D}_\a$ 
to the right, there appear contributions proportional to  space-time derivatives, due to the identity
\bea
\{ {\mathfrak D}_\a , {\bar \nabla}_{\dt \b}  \}=
-2{\rm i}\, 
(\s^m )_{\a \dt \b} \,\pa_m ~,
\eea
 which
do not contribute to the action. This completes the proof.

An important property of the action (\ref{PAP}) is its
invariance under arbitrary {\it projective transformations} of the form:
\be
(u_i \,,\,v_i )~\to~(u_i\,,\, v_i )\,R~,~~~~~~R\,=\,
\left(\begin{array}{cc}a(t)~&0\\ b(t)~&c(t) \end{array}\right)\,\in\,{\rm GL(2,\mathbb{C})}~,
\label{projectiveGaugeVar}
\ee
where the matrix elements of $R$ obey the first-order equations
\bea
{\dt a} = b \frac{({\dt v},v)}{(v,u)}~, \qquad 
{\dt b} = -b \frac{({\dt v},u)}{(v,u)}~, 
\qquad {\dt \j}:= \frac{{\rm d} \j(t)}{\rm d t} 
\eea
along the integration contour in order to keep  the transformed isotwistor $u_i$  $t$-independent.  
This invariance allows one to make $u_i$ arbitrary modulo the constraint 
$(v,u)\neq 0$, and therefore the action is independent of $u_i$, 
\bea
\frac{\pa }{\pa u_i} { S}=0~.
\eea

The projective-superspace action was originally given in \cite{KLR} in a form that differs slightly 
from (\ref{PAP}).  The latter representation appeared first in  \cite{Siegel85}.

\section{Off-shell projective supermultiplets}
\setcounter{equation}{0}

We now turn to a systematic study of projective supermultiplets.

\subsection{Projective superfields in the north chart  of ${\mathbb C}P^1$}
Introduce the inhomogeneous complex coordinate, $ \z$, on  ${\mathbb C}P^1 
-\{\infty\}$
defined by eq. (\ref{Zeta}).
Given a weight-$n$ projective superfield $ Q^{(n)}(z,v)$, we can associate with it 
a new object $ Q^{[n]}(z,\z )$ defined as 
\bea
Q^{(n)}(z,v)~\longrightarrow ~ Q^{[n]}(z,\z) \propto Q^{(n)}(z,v)~, \qquad 
\frac{\pa}{\pa \bar \z} Q^{[n]} =0 ~.
\eea
The explicit form of  $ Q^{[n]}(z,\z) $ for various projective multiplets will be given later on.
The superfield introduced  can be represented by a  series
\bea
Q^{[n]}(z,\z) =
\sum_{p}^{q} Q_k (z) \z^k~,\qquad -\infty \leq p <q \leq +\infty~,
\label{seriess}
\eea
with $Q_k(z)$ some {ordinary $\cN=2$ superfields}. Here $p$
and $q$ are {\it invariants} of the supersymmetry transformations.

In the north chart  of ${\mathbb C}P^1$, the analyticity constraints
\bea
 {\mathfrak D}_{ \a} Q^{(n)} = {\bar {\mathfrak D}}_{\dt \a } Q^{(n)} =0~, \qquad
{ {\mathfrak D}_{ \a} :=v_i \,{ D}^i_{ \a} ~, \quad 
{\bar {\mathfrak D}}_{\dt \a} := v_i \,{\bar { D}}^i_{\dt \a}}
\label{AnaCon2}
\eea
take the form: 
\bea
{ D^{\2}_{\a}}
Q^{[n]}(\z)=\z\,{ D^{\1}_{\a}}Q^{[n]}(\z)~, \qquad 
{ {\bar D}_{  {\dt \a}\, \2}}Q^{[n]}(\z)=-\frac{1}{\z}\,
{ {\bar D}_{ {\dt \a}\, \1}}Q^{[n]}(\z)~.
\label{ancon}
\eea
These relations can be interpreted  as follows.
The dependence of the component superfields 
$Q_k $ of $Q^{[n]}(\z)$
on $\q^\a_{\2}$ and ${\bar \q}^{\2}_{\dt \a}$,
is uniquely determined in terms 
of their dependence 
on the variables $\q^\a_{\1}\equiv \q^\a$
and ${\bar \q}^{\1}_{\dt \a}\equiv {\bar \q}_{\dt \a}$,
which can be identified with the 
Grassmann coordinates of $\cN=1$ superspace parametrized by 
$z^M= (x^m, \q^\a ,{\bar \q}_{\dt \a}$).

\subsection{Smile conjugation}
\label{smile}
The notion of smile conjugation was introduced in subsection \ref{Iso-superspace}. 
As formulated, the  definition
 directly applies to $Q^{(n)}(z,v)$. Now we wish to re-express it
in terms of   $Q^{[n]}(z,\z)$. 

Consider a projective superfield\footnote{As compared with (\ref{seriess}), 
we have changed $p \to -p$ in eq. (\ref{seriess2}).}
\bea
Q (z,\z) \equiv Q^{[n]}(z,\z) =
\sum_{-p}^{q} Q_k (z) \z^k~.
\label{seriess2}
\eea
It is constrained as in eq. (\ref{ancon}).
Let ${\bar Q}(z, \bar \z)$ be 
the {complex conjugate} of $Q(z,\z)$, 
\bea
{\bar Q} (z, \bar \z) =
\sum_{-p}^{q} {\bar Q}_k (z) {\bar \z}^k~,
\qquad  {\bar Q}_k (z) := \overline{ Q_k (z) }~.
\eea
It is not a projective superfield, for it satisfies the conditions
\bea
 D^{\2}_{\a} 
{\bar Q}  ( \bar \z)= -\frac{1}{\bar \z} \,{ D^{\1}_{\a}}{\bar Q} ( \bar \z)~, \qquad 
{ {\bar D}_{  {\dt \a}\, \2}} \bar Q (\bar  \z)= {\bar \z}\,
{ {\bar D}_{ {\dt \a}\, \1}} \bar Q (\bar \z)~,
\eea
which do not coincide with the analyticity constraints.
However, the following object
\bea
\breve{Q} (z, \z ) 
:=  {\bar Q} \left (z,  -\frac{1}{\z} \right)
=\sum_{-q}^{+p} (-1)^k{\bar Q}_{-k} (z) { \z^k}~
\eea
does obey the analyticity constraints, and therefore it is a projective superfield.
The $\breve{Q} (\z)$ is called the smile-conjugate of $Q(\z)$.

A real projective superfield is characterized by the properties:
\bea
 \breve{Q} (z,\z) =Q(z, \z) =
{ \sum_{-p }^{+p} Q_k (z) \z^k}~, \qquad 
{\bar Q}_k(z) = (-1)^k Q_{-k}(z).
\label{realPS}
\eea

\subsection{$\cN = 2$ supersymmetric action in $\cN = 1$ superspace}
Consider the $\cN=2$ supersymmetric action
\bea
S:=- \frac{1}{2\p} \oint_\g  { v_i {\rm d} v^i }
\int {\rm d}^4x \, \D^{(-4)} \cL^{(2)} (z,v) \Big|_{\q_i={\bar \q}^i =0} ~.
\eea
We recall that  $\cL^{(2)}(z,v)$ is a real weight-2 projective superfield, 
\bea
\D^{(-4)} := \frac{1}{16}\nabla^2
{\bar \nabla}^2
~, \qquad
\nabla_\a := \frac{1}{(v,u)} {u_i} D^i_\a ~, \quad 
{\bar \nabla}_{\dt \b} := \frac{1}{(v,u) } {u_i} {\bar D}^i_{\dt \b} ~,
\eea
and $u_i$ is a {fixed} isotwistor such that $(v,u) \neq 0$ at each point of  $ \g$.
As demonstrated in subsection \ref{PSA}, the action is independent of $u_i$.

Without loss of generality,  we can assume  that the integration contour $\g$
does not pass through the ``{north pole}'' $v^{i} \sim (0,1)$.
Then, we can introduce the inhomogeneous complex coordinate, $ \z$, 
on  ${\mathbb C}P^1 -\{\infty\}$ defined by $ v^i = v^{\1} \,(1, \z)$.
Since the action, $S$, is independent of $u_i$, the latter can be chosen to be   $ u_i =(1,0)$, 
such that $(v,u) = v^{\1}\neq 0$. 
We also represent the Lagrangian in the form: 
\bea
\cL^{(2)}(z,v)={\rm i} \,v^{\1}v^{\2}\cL(z,\z)
= {\rm i} (v^{\1})^2 \,\z\,{ \cL(z,\z)~, 
\qquad \breve{\cL} =\cL}~. 
\eea
It is important to remark that $\cL(z,\z)$ is a real projective superfield in the sense of eq. (\ref{realPS}).
Now, the action takes the form:
\bea
S= \frac{1}{ 16}\oint  \frac{\rd\z }{ 2\pi\ri}
\int\rd^4 x\,\z\,
({D}^{\1})^2({\bar D}_{\2})^2\cL(z,\z)\Big|_{\q_i={\bar \q}^i =0} ~.
\label{ac2}
\eea
${}$Finally, if we make use of the analyticity of $\cL$, 
\bea
D^{\2}_{\a}\cL(\z)=\z\,D^{\1}_{\a}\cL(\z)~, \qquad 
{\bar D}_{\2}^{\dt \a} \cL(\z)=-\frac{1}{\z}\,{\bar D}_{\1}^{\dt \a}\cL(\z)~,
\eea
the action turns into 
\bea
S &=& \frac{1 }{ 2\pi\ri }
 \oint 
 \frac{\rd\z }{  \z}
\int\rd^4 x\, \Big\{ \frac{1}{16} ({D}^{\1})^2({\bar D}_{\1})^2\Big\}
\cL(z,\z)\Big|_{\q_i={\bar \q}^i =0} \non \\
&=& \frac{1 }{ 2\pi\ri }
 \oint 
 \frac{\rd\z }{  \z}
\int\rd^4 x\,{\rm d}^4\q \,
\cL(z,\z)\Big|_{\q_{\2}={\bar \q}^{\2} =0}~.
\label{6.15}
\eea
In the final expression for $S$, 
the integration is carried out over the $\cN=1$ superspace.\footnote{In what follows, 
the bar-projection in expressions like the second line in (\ref{6.15}) is omitted.} 
The action is now formulated entirely in terms of $\cN=1$ superfields. 
At the same time, by construction, it is off-shell $\cN=2$ supersymmetric! 
This is one of the most powerful features of the projective superspace approach.

\subsection{Projective multiplets and constrained $\cN = 1$ superfields} 
There is an important feature of projective multiplets that has to be specially emphasized.
Consider a projective multiplet
\bea
Q^{[n]}(z,\z) =
\sum_{p}^{q} Q_k (z) \z^k~,\qquad { -\infty \leq p <q \leq +\infty} ~.
\label{6.16}
\eea
In terms of  $Q_k$, the analyticity conditions are:
\bea
D^{\2}_{\a}Q_k=D^{\1}_{\a}Q_{k-1}~, \qquad 
{\bar D}_{\2}^{\dt \a}Q_{k-1}=-{\bar D}_{\1}^{\dt \a}Q_k~.
\eea

Suppose that the series (\ref{6.16}) terminates from below, that is $ p>-\infty$.
Then $Q_p$ and $Q_{p+1} $ can be seen to be  constrained $\cN=1$ superfields. 
The corresponding constraints are:
\bea
{\bar D}^{\dt \a} Q_p = 0~, \qquad {\bar D}^2 Q_{p+1} =0 ~,
\qquad 
 { {\bar D}^{\dt \a}:= {\bar D}_{\1}^{\dt \a}}~.
\eea
Thus $Q_p$ is {\it chiral}, while $Q_{p+1}$ is said to be {\it linear}.

Suppose the series terminates from above, that is $q < \infty$.
Then, the  $\cN=1$ superfields $Q_q$ and $Q_{q-1} $ are constrained by 
\bea
{D}_{ \a} Q_q = 0~, \qquad { D}^2 Q_{q-1} =0 ~,
 \qquad  D_\a :=D^{\1}_{\a}~.
\eea
Thus $Q_q$ is {\it antichiral}, while $Q_{q-1}$ is said to be {\it antilinear}.

There is a very special case: $q-p=2$.  Here the $\cN=1$ superfield components 
are constrained by the rule:
\bea
{\bar D}_{\dt \a} Q_p = 0~, \qquad {\bar D}^2 Q_{p+1} = D^2 Q_{p+1} =0 ~, 
\qquad {D}_{ \a} Q_{p+2} = 0~.
\eea
We see that $Q_{p+1} $ is both linear and antilinear.

\subsection{Off-shell realizations of the hypermultiplet}
We now review off-shell projective multiplets that can be used to describe
the $\cN=2$ scalar multiplet, also known as the {\it hypermultiplet}, comprising four spin-0 
and two spin-1/2 fields. The $\cN=2$ supersymmetric nonlinear sigma-models can be viewed
as models for self-interacting massless hypermultiplets.

Our first example is the so-called real $\cO(2n)$ multiplet \cite{KLT,LR-projective1}, $n=2,3\dots$, which 
 is described by a real weight-$2n$ projective superfield $H^{(2n)} (z,v) $ of the form:
\bea
H^{(2n)} (z,v) &=& H_{i_1 \dots i_{2n}}(z) v^{i_1} \dots v^{i_{2n}} 
=\breve{H}^{(2n)} (z,v) ~.
\eea
The analyticity constraints (\ref{AnaCon2}) are equivalent to 
\bea
D_{\a (j} H_{i_1 \dots i_{2n} )} ={\bar D}_{\dt \a (j} H_{i_1 \dots i_{2n} )} =0~.
\eea
The reality condition $\breve{H}^{(2n)}  = {H}^{(2n)} $ is equivalent to 
\bea
\overline{ H_{i_1 \dots i_{2n}} } &=& H^{i_1 \dots i_{2n}}
=\ve^{i_1 j_1} \cdots \ve^{i_{2n} j_{2n} } H_{j_1 \dots j_{2n}} ~.
\eea
Associated with $H^{(2n)} (z,v) $ is the superfield $H^{[2n]}(z,\z) $ defined by 
\bea
H^{(2n)} (z,v) &=&\big({\rm i}\, v^{\1} v^{\2}\big)^n H^{[2n]}(z,\z) =
\big(v^{\1}\big)^{2n} \big({\rm i}\, \z\big)^n H^{[2n]}(z,\z)~,  \non \\
H^{[2n]}(z,\z) &=& 
\sum_{k=-n}^{n} H_k (z) \z^k~,
\qquad  {\bar H}_k = (-1)^k H_{-k} ~.
\label{o2n1}
\eea
The $H^{[2n]}(z,\z)$  is real in the sense of (\ref{realPS}). Its two lowest components
in the expansion (\ref{o2n1}),
$H_{- n} $ and $H_{-n+1}$, are constrained $\cN=1$ superfields, chiral and linear, respectively, 
\bea
 {\bar D}_{\dt \a} H_{-n} &=&0~, 
\qquad {\bar D}^2 H_{-n+1} =0~. 
\label{o2n11}
\eea

In the family of multiplets considered above, 
we intentionally did not include the real $\cO(2)$ multiplet \cite{KLR} 
described by 
\bea
\eta(z, \z) = \frac{1}{\z}\, \vf (z) + G(z) - \z \,{\bar \vf}(z)~, \quad  {\bar G}=G~, \quad
{\bar D}_{\dt \a} \vf = {\bar D}^2 G =0~.
\label{tensor-series}
\eea
The point is that this multiplet is very special, for it
corresponds to the $\cN=2$ {\it tensor multiplet} \cite{Wess} in which one of the four spin-0 states
is described by a gauge antisymmetric second rank tensor field. 
 
All of the $\cO(2n)$ multiplets, with $n=1,2,\cdots $,  prove to define holomorphic tensor fields over
${\mathbb C}P^1$. We now turn to introducing projective multiplets that are not globally 
defined on ${\mathbb C}P^1$. By definition, the arctic multiplet  \cite{LR-projective1} 
is described by a series
\bea
\U (z, \z) = \sum_{k=0}^{\infty} \U_k (z) \z^k~, \qquad 
{\bar D}_{\dt \a} \U_0 =0~, 
\qquad {\bar D}^2 \U_1 =0~ .
\label{6.27}
\eea
Its smile-conjugate, $\breve{\U}(z, \z) $, is called an antarctic multiplet,
\bea
\breve{\U}(z, \z) = \sum_{k=0}^{\infty} (-1)^k {\bar \U}_k (z)
\frac{1}{\z^k}~.
\label{6.28}
\eea
The superfields  ${\U}(z, \z)$ and $\breve{\U}(z, \z) $ constitute a polar multiplet.
This terminology, {\it (ant)arctic} and {\it polar},
was coined in  \cite{G-RRWLvU} and appears to be quite natural, since
several practitioners  of projective superspace come from the Nordic country of Sweden.

Among the projective multiplets considered, the polar multiplet has two unique properties.
First of all, it is the only multiplet which can be used to describe a charged hypermultiplet, since the 
structure of the arctic multiplet allows for phase transformations 
\bea
\U(\z) ~\longrightarrow ~ {\rm e}^{{\rm i} \a } \U(\z) ~, \qquad \a \in {\mathbb R}~.
\eea
Second, the space of arctic superfields allows for a ring structure: 
for any arctic superfields $\U_A (\z)$ and $\U_B  (\z)$, their product 
\bea
\U_A(\z)  \cdot \U_B (\z) =\U_C(\z)
\eea
is also arctic.

\section{Sigma-models in projective superspace}
\setcounter{equation}{0}

We are finally prepared to write down general off-shell $\cN=2$ supersymmetric nonlinear sigma-models.

\subsection{General off-shell $\cN = 2$ supersymmetric sigma-models}
Suppose we have a dynamical system described by a set of $\cN=2$ tensor multiplets.
Then, their most general $\cN=2$ supersymmetric sigma-model couplings 
are realized by actions of the form \cite{KLR,GHR}:
\bea
S_{\rm tensor}= \frac{1 }{ 2\pi\ri }
 \oint 
 \frac{\rd\z }{  \z}
\int\rd^4 x\,{\rm d}^4\q \,
\cL\Big(\eta {(\z)} ;\z \Big)~,
\label{SM-tensor}
\eea
with $\eta (\z) $ given by eq.  (\ref{tensor-series}).
Upon evaluation of the contour integral, the action can be shown to reduce to that 
constructed originally in the $\cN=1$ superspace setting in \cite{LR}.\footnote{Incidentally, 
using
the general results on self-interacting $\cN=2$ tensor multiplets obtained in \cite{LR},  
the representation (\ref{SM-tensor}) 
could have been discovered already in 1983, if the authors of \cite{LR} had used
a classical formula of Whittaker for harmonic functions in ${\mathbb R}^3$ \cite{Whittaker}.}

Similarly, in the case of  $\cO (2n)$ multiplets defined by eqs. (\ref{o2n1}) and (\ref{o2n11}), their 
general $\cN=2$ supersymmetric sigma-model couplings are described by actions of the form
\cite{KLT,LR-projective1}:
\bea
S_{\cO  }= \frac{1 }{ 2\pi\ri }
 \oint 
 \frac{\rd\z }{  \z}
\int\rd^4 x\,{\rm d}^4\q \,
\cL\Big(H^{[\dots ]} {(\z)};\z \Big)~.
\label{SM-O(2n)}
\eea

In the case of  polar multiplets defined by eqs. (\ref{6.27}) and (\ref{6.28}), 
their general sigma-model couplings are 
described by actions of the form
\cite{LR-projective1}:
\bea
S_{\rm polar}= \frac{1 }{ 2\pi\ri }
 \oint 
 \frac{\rd\z }{  \z}
\int\rd^4 x\,{\rm d}^4\q \,
\cL\Big(\U  { (\z)}, \breve{\U}  (\z) ;\z \Big)~.
\label{SM-polar}
\eea

${}$Finally, the most general off-shell  $\cN=2$ supersymmetric sigma-models 
describe couplings 
of tensor multiplets, $\cO(2n)$ multiplets and polar multiplets.
\bea
S_{\rm general}= \frac{1 }{ 2\pi\ri }
 \oint 
 \frac{\rd\z }{  \z}
\int\rd^4 x\,{\rm d}^4\q \,
\cL\Big(\eta  (\z), H^{[\dots ]}  (\z) , \U  (\z), \breve{\U}  (\z); 
\z \Big)~.
\label{SM-genaral}
\eea

In all of the off-shell $\cN=2$ supersymmetric sigma-models introduced, 
the Lagrangian may depend explicitly on $\z$. Each of  the Lagrangians 
\bea
 \cL\big(\eta;\z \big)~,  \qquad \cL\big(H^{[\dots ]};\z \big)~, \qquad \cL\big(\U, \breve{\U};\z \big)
 ~, \qquad \cL\big(\eta, H^{[\dots ]} , \U, \breve{\U}; 
\z \big)
\non 
\eea
should be an analytic function of its arguments, but otherwise arbitrary,
modulo a reality condition  with respect  to the  smile conjugation.

\subsection{Generalized Legendre transform construction}
\label{GeneralizedLT}
The action (\ref{SM-genaral}) provides us with the most general off-shell $\cN=2$ supersymmetric 
sigma-models that can be constructed in projective superspace. The Lagrangian in (\ref{SM-genaral}) 
can be chosen at will, modulo mild restrictions discussed earlier. Different choices of 
the Lagrangian will lead, in general, to different hyperk\"ahler metrics in target space. 
So, it  is natural to ask: Does projective superspace offer us a free lunch?
In other words, can we immediately read off the target space metric from (\ref{SM-genaral})? 
The answer is ``No'' in general. Except in the very special case 
of  tensor models (\ref{SM-tensor}), which will be discussed separately, 
one has to go through a  technical procedure known as the 
{\it generalized Legendre transform construction},  originally sketched in \cite{LR-projective1},
in order to derive a  hyperk\"ahler metric from (\ref{SM-genaral}).
It is called `generalized' because it is an extension of the so-called {\it linear Legendre transform construction}
\cite{LR,KLR,HitchinKLR} to be discussed in the next subsection.

To fix the ideas, consider a $\cN=2$ supersymmetric nonlinear sigma-model described
either by a single {$\cO(2n)$ multiplet} ($n\geq 2$) or by a polar multiplet.
Upon evaluation of the contour integral, the action becomes
\bea
S =  \int  \rd^4 x\,{\rm d}^4\q\, 
L_{\text{off-shell}}(\F , \bar \F, \S , \bar \S, \cU_\imath)~,
\label{af1}
\eea
for some Lagrangian $L_{\text{off-shell}}(\F , \bar \F, \S , \bar \S, \cU_\imath)$.
The dynamical variables of the theory consist of
(i) two {\it physical} superfields $\F$ and $\S$ and their conjugates $\bar \F$ and $\bar \S$;
and (ii) some number of {\it auxiliary} superfields
$\cU_\imath$.
Here the index $\imath$ may take a finite ($2n-3$, in the case of $\cO(2n)$ multiplet)
or infinite (in the case of  polar multiplet) number of values.
The physical superfields $\F$ and $\S$ are chiral and complex linear,
\bea
 {\bar D}_{\dt{\a}} \F =0~, \qquad \qquad {\bar D}^2 \S = 0 ~,
\label{chiral+linear}
\eea
while the auxiliary superfields $\cU_\imath$ are {\it unconstrained}.
The $\cU$s are auxiliary, for their Euler-Lagrange equations are algebraic
\bea
\frac{\pa }{\pa \cU_{\jmath} }L_{\text{off-shell}}(\F , \bar \F, \S , \bar \S, \cU_\imath)=0~.
\eea
Under reasonable regularity conditions on the Lagrangian, these equations uniquely determine the auxiliary
superfields as functions of the physical ones, 
 \bea
 \cU_\imath = \cU_\imath (\F , \bar \F, \S , \bar \S)~.
\eea
This leads to  an action formulated in terms of the physical superfields: 
\bea
 S &=&  \int  \rd^4 x\,{\rm d}^4\q\, 
L(\F , \bar \F, \S , \bar \S)~,
\label{af3}
\\
L(\F , \bar \F, \S , \bar \S)&:=& 
L_{\text{off-shell}}\Big(\F , \bar \F, \S , \bar \S, \cU_\imath (\F , \bar \F, \S , \bar \S)\Big)~.
\non 
\eea
This action is of course $\cN=2$ supersymmetric, 
however only one of the two supersymmetries is  manifest. 
Since the auxiliaries have been eliminated, the first and second supersymmetry transformations
form the $\cN=2$ super-Poincar\'e algebra only on the mass shell.

Even though  the action (\ref{af3})  is formulated in terms of the physical superfields only, 
the Lagrangian $L(\F , \bar \F, \S , \bar \S)$
is not a hyperk\"ahler potential, since  the dynamical variable $\S$ is {complex linear}. 
As discussed in subsection \ref{N=2-->N=1},  the Lagrangian   coincides with 
the  hyperk\"ahler potential of the target space provided the theory is formulated 
in terms of {chiral} superfields and their conjugates only.
Is it possible to develop such a (re)formulation for  the theory (\ref{af3})? 
The answer is affirmative indeed  under reasonably  general conditions,  
due to the existence of a duality between chiral and complex
linear superfields that was  noticed for the first time by Zumino \cite{Zumino1980}.

It has been known for thirty years \cite{GS}
that the chiral and complex linear superfields 
provide different off-shell descriptions of the free $\cN=1$ scalar multiplet, which 
are known as the minimal and non-minimal scalar multiplet models, respectively.
They are described by the following actions:
\begin{subequations}
\bea
S_{\rm minimal}&=& \phantom{-}\int \rd^4 x\,{\rm d}^4\q \, {\bar \J}\, \J ~, \qquad {\bar D}_{\dt \a} \J =0~,
\label{Min-scalar}\\
S_{\text{non-minimal}} &=& -\int \rd^4 x\,{\rm d}^4\q \, {\bar \S} \,\S ~, \qquad {\bar D}^2 \S=0~.
\label{Non-Min-scalar}
\eea
\end{subequations}
It is easy to read off  the corresponding equations of motion.\footnote{In deriving the equations of motion 
for $\J$ and $\S$, it is useful to represent $\J = {\bar D}^2 \bar R$ and $\S= {\bar D}_{\dt \a} {\bar \x}^{\dt \a}$, 
for unconstrained superfields $\bar R$ and ${\bar \x}^{\dt \a}$.} 
On the mass shell, the dynamical superfields must obey the off-shell constraints and the equations of motion.
It is convenient to combine them in a simple table: 
\bea
\begin{array}{|c|c|c|}
\hline
\phantom{\Big|}\mbox{free scalar multiplet} ~ {} & ~\mbox{off-shell constraint} ~& ~\mbox{equation of motion} \\
  \hline
\phantom{\Big|}\mbox{minimal} &{ {\bar D}_{\dt \a} \J =0} &{  D^2 \J =0} \\
\hline
\phantom{\Big|}\ \mbox{non-minimal} &{  D^2 {\bar \S} =0} & {  {\bar D}_{\dt \a} \bar \S=0} \\
\hline
\end{array}
\qquad {}
 \non
\eea
One can see that the two models (\ref{Min-scalar}) and (\ref{Non-Min-scalar})
are dynamically equivalent.
Moreover, these models are dual to each other. This means that they are related to each other
through the use of a first-order action.
Such an action can be chosen  \cite{Zumino1980} to be
\bea
S_{\text{first-order}}=   \int \rd^4 x\,{\rm d}^4\q \, 
\Big\{- \bar \G \,\G 
+\J \,\G + {\bar \J} {\bar \G} 
\Big\}~.
\label{f-o1}
\eea
Here  $\G$ is {complex unconstrained}, 
while $\Psi$ is {chiral}, ${\bar D}_{\dt \a} \J =0$.
Varying this action with respect  to $\J$ gives $\G= \S$, 
and then $S_{\text{first-order}}$ reduces to  (\ref{Non-Min-scalar}). 
On the other hand, the equation of motion for $\G$ implies $\bar \G = \J$, 
and then $S_{\text{first-order}}$ reduces to  (\ref{Min-scalar}). 

Let us generalize the simple example analyzed above. Consider a theory of self-interacting 
complex linear superfields $\S^a$ and their conjugates ${\bar \S}^{\bar a}$ described 
by an action of the form
\bea
S=\int \rd^4 x\,{\rm d}^4\q \, \cL ( \S^a , {\bar \S}^{\bar b}) ~, 
\label{NM-sigma}
\eea
where the Lagrangain $\cL ( \S, {\bar \S})$ is  
a real analytic function of the dynamical superfields.
By analogy with (\ref{f-o1}), we can associate with (\ref{NM-sigma}) 
the following first-order action:
\bea
S_{\text{first-order}}=   \int \rd^4 x\,{\rm d}^4\q \, 
\Big\{ 
\cL (\G , \bar \G) 
+\J_a \,\G^a + {\bar \J}_{\bar a} {\bar \G}^{\bar a} 
\Big\}~,
\label{f-o2}
\eea
where $\G^a$ are {complex unconstrained}, 
while $\Psi_a$  {chiral}, ${\bar D}_{\dt \a} \J_a =0$.
This theory is equivalent  to  (\ref{NM-sigma}). Indeed,  varying (\ref{f-o2} with respect to 
$\J_a$ gives $\G^a =\S^a$, and then (\ref{f-o2}) reduces to  (\ref{NM-sigma}). 
Now, consider the equations  of motion for $\G^a$ and ${\bar \G}^{\bar a} $:
\bea
\frac{\pa  }{\pa \G^a } \cL(\G, \bar \G) + \J_a =0~, 
\qquad \frac{\pa  }{\pa {\bar \G}^{\bar a} } \cL(\G, \bar \G) + {\bar \J}_{\bar a} =0~. 
\label{EoM-Gamma}
\eea
These equations allow one to express $\G$s and $\bar \G$s in terms of $\J$s and $\bar \J$s
provided
\bea
\det  \left(
\begin{array}{c  c}
 \cL_{a b}    ~& ~  \cL_{a \bar b}   \\
\cL_{\bar a b} 
& \cL_{\bar a \bar b}   \\
\end{array}
\right) \neq 0~.
\label{Hessian1}
\eea
Then, the action (\ref{f-o2}) turns into
\bea
S_{\text{dual}}&=&   \int \rd^4 x\,{\rm d}^4\q \, 
\cK(\J_a, {\bar \J}_{\bar b})~,
 \label{dual} 
 \eea
 where we have defined 
\bea
\cK(\J, \bar \J) &:=&
\Big\{ \cL (\G , \bar \G) 
+\J_a \,\G^a + {\bar \J}_{\bar a} {\bar \G}^{\bar a} 
\Big\}\Big|_{\G = \G(\J, \bar \J)}~.
\label{LT-K}
\eea
It is clear that $\cK(\J, \bar \J)$ is  (up to a trivial sign difference) 
the Legendre transform of $\cL(\G, \bar \G)$.
Standard properties of the Legendre transformation now imply 
\bea
\frac{\pa  }{\pa \J_a } \cK(\J, \bar \J) - \G^a =0~, 
\qquad \frac{\pa  }{\pa {\bar \J}_{\bar a} } \cK(\J, \bar \J)-  {\bar \G}^{\bar a} =0~ 
\label{EoM-Psi}
\eea
as well as
\bea
\det  \left(
\begin{array}{c  c}
 \cK^{a b}    ~& ~  \cK^{a \bar b}   \\
\cK^{\bar a b} 
& \cK^{\bar a \bar b}   \\
\end{array}
\right) \neq 0~.
\label{Hessian2}
\eea

It is natural to interpret the Lagrangian in (\ref{dual}) as  the K\"ahler potential of a K\"ahler 
manifold. For such an  interpretation to be consistent, 
it must hold that 
\bea
\det \left( \frac{\pa \cK}{\pa \J_a {\bar \J}_{\bar b}} \right)
 \neq 0~.
\label{Hessian3}
\eea
Then, due to (\ref{Hessian1}) and (\ref{Hessian2}), we also must have 
\bea
\det \left( \frac{\pa \cL}{\pa \S^a {\bar \S}^{\bar b}} \right)
\neq 0~.
\label{Hessian4}
\eea
The latter condition is equivalent to the fact that, say,  the first equation in (\ref{EoM-Gamma})
can be solved to express the variables $\bar \G$s as functions of $\J$s and $\G$s.

Our consideration shows that the requirements (\ref{Hessian1}) and (\ref{Hessian4})
are essential for the theory (\ref{NM-sigma}) to provide a dual description of  
$\cN=1$ supersymmetric nonlinear sigma-models.

Before returning to the theory of our interest, eq. (\ref{af3}), it is worth mentioning 
another important aspect concerning the dual theories (\ref{NM-sigma}) and (\ref{dual}).
One can develop a dual version of (\ref{dual}) by considering a first-order action of the form:
\bea
S_{\text{first-order}}=   \int \rd^4 x\,{\rm d}^4\q \, 
\Big\{ 
\cK (U , \bar U) 
-\S^a  U_a - {\bar \S}^{\bar a} {\bar U}_{\bar a} 
\Big\}~,
\label{f-o3}
\eea
where the superfields $U_a$ are complex unconstrained, and $\S^a$ complex linear.
The variables $U$s and $\bar U$s can be integrated out, due to (\ref{Hessian2}).
If $\cK$ coincides with (\ref{LT-K}), one then ends up with (\ref{NM-sigma}). 
However, the Lagrangian in  (\ref{dual}) is defined modulo K\"ahler transformations
\bea
\cK(\J, \bar \J) ~~\longrightarrow  ~~ \widetilde{\cK} (\J, \bar \J) = \cK(\J, \bar \J) +\L (\J) +{\bar \L} (\bar \J )~,
\eea
with $\L(\J)$ an arbitrary holomorphic function. If one replaces $\cK \to \widetilde{\cK} $ 
in (\ref{f-o3}), and then integrates out the variables $U$s and $\bar U$s, 
the resulting theory will be described by a Lagrangian $\widetilde{ \cL} ( \S^a , {\bar \S}^{\bar b}) $
that differs from that appearing in  (\ref{NM-sigma}). 
Actually, applying K\"ahler transformations  may lead to quite a bizarre situation. 
The point is that the transformed K\"ahler potential,  $\widetilde{\cK} (\J, \bar \J)$, 
always obeys the inequality (\ref{Hessian3}). However, eq. (\ref{Hessian2})  
may not hold\footnote{As an  example, consider
$ {\cK} (\J, \bar \J) = \bar \J \J $ and choose $\widetilde{\cK} (\J, \bar \J) =  \bar \J \J  +(\a /2) (\J^2 + {\bar \J}^2)$, 
with $\a $ a constant parameter.  Eq.  (\ref{Hessian2})  does not hold for  $\widetilde \cK$ if $\a = \pm 1$.} 
for $\widetilde \cK$, and then the procedure of integrating out the variables $U$s and $\bar U$s 
from (\ref{f-o3}) becomes more involved.

${}$Finally, let us return to our  sigma-model  (\ref{af3}). It  
is equivalent to the following first-order action:
\bea
S_{\text{first-order}}=   \int \rd^4 x\,{\rm d}^4\q \, 
\Big\{\,
L (\F , \bar \F, \G , \bar \G)~
+\J \,\G + {\bar \J} {\bar \G} 
\Big\}~.
\eea
Integrating out $\G$ and $\bar \G$ leads to an action of the form
\bea
S_{\rm dual}=   \int \rd^4 x\,{\rm d}^4\q \, 
H(\F , \bar \F, \J , \bar \J)~,
\eea
where $H(\F , \bar \F, \J , \bar \J)$  is the Legendre transform of
$L (\F , \bar \F, \S , \bar \S)$, with $\F$ and $\bar \F$ being treated as parameters.
The resulting Lagrangian, 
$H(\F , \bar \F, \J , \bar \J)$, 
is the K\"ahler potential of a hyperk\"ahler manifold.

\subsection{Linear Legendre transform construction}
Here we briefly review the famous {\it linear Legendre transform construction}
\cite{LR,KLR,HitchinKLR}. Our presentation is intentionally brief, for it is hardly possible 
to present this construction better than it has already been done in \cite{HitchinKLR}.

The most  general $\cN=2$  supersymmetric sigma-model coupling of  several tensor multiplets $\eta^i(\z)$
is described by an action of the form \cite{KLR}:
\bea
S= \frac{1 }{ 2\pi\ri }
 \oint 
 \frac{\rd\z }{  \z}
\int\rd^4 x\,{\rm d}^4\q \,
\cL\Big(\eta^i {(\z)} ;\z \Big)~,
\label{TensorAction}
\eea
where the dynamical variables are:
\bea
\eta^i (\z) = \frac{1}{\z}\, \vf^i + G^i - \z \,{\bar \vf}^i~, \qquad {\bar D}_{\dt \a} \vf^i =0~, 
\qquad {\bar D}^2 G^i ={\bar G}^i-G^i=0~.
\eea
Unlike the multiplets considered in the previous subsection, 
the tensor multiplet  requires no $\cN=1$ auxiliary superfields.
Suppose we have  evaluated the contour integral in (\ref{TensorAction}). Then, 
 the action turns into
\bea
S=  
\int\rd^4 x\,{\rm d}^4\q \,
L\big(\vf^i , {\bar \vf}^i , G^i \big)~. 
\label{TensorAction2}
\eea
Here the Lagrangian cannot yet be identified with a hyperk\"ahler potential, 
for the superfields $G^i$ are real linear.
In order to derive the hyperk\"ahler potential of the target space,  
we have to dualize each $\cN=1$ tensor multiplet, 
$G^i$,  into  a chiral superfield  $\J_i$ and its conjugates ${\bar \J}_i$.
It is worth studying in some more detail how such a duality works.

The $\cN=1$ tensor multiplet \cite{Siegel} provides a variant off-shell realization of 
the massless scalar multiplet in which one of the two scalar fields is dualized into a gauge 
antisymmetric tensor field. Consider the models for free chiral and real linear superfields:
\begin{subequations}
\bea
S_{\rm scalar}&=& \hf \int \rd^4 x\,{\rm d}^4\q \, (\J +{\bar \J})^2 ~, \qquad {\bar D}_{\dt \a} \J =0~,
\label{scalar}\\
S_{\text{tensor}} &=& -\hf\int \rd^4 x\,{\rm d}^4\q \, G^2 ~, \qquad 
{\bar D}^2 G ={\bar G}-G=0
~.
\label{tensor}
\eea
\end{subequations}
Here the action (\ref{scalar}) can be seen to coincide with (\ref{Min-scalar}).
The constraint  on $G$ is solved  \cite{Siegel}
by introducing a chiral spinor prepotential $\eta_\a$,  by the  rule 
\bea
G= D^\a \eta_\a +{\bar D}_{\dt \a} {\bar \eta}^{\dt \a} ~, \qquad {\bar D}_{\dt \a} \eta_{\b} =0~.
\label{G1}
\eea
The prepotential  is defined modulo gauge transformations of the form:
\bea
\d \eta_\a = {\rm i}\, {\bar D}^2 D_\a V ~, \qquad V = \bar V.
\label{G2}
\eea
The theories (\ref{scalar}) and (\ref{tensor}) are dynamically equivalent, 
as can be seen from the following table:
\bea
\begin{array}{|c|c|c|}
\hline
\phantom{\Big|} ~ {} & ~\mbox{off-shell constraint} ~& ~\mbox{equation of motion} \\
  \hline
\phantom{\Big|}\mbox{scalar multiplet} ~&{ {\bar D}^2 D_{\a} (\J +{\bar \J})=0} &{  {\bar D}^2 (\J +{\bar \J}) =0} \\
\hline
\phantom{\Big|} \mbox{tensor multiplet} ~&{ {\bar  D}^2 G =0} & {  {\bar D}^2 D_{ \a} G=0} \\
\hline
\end{array}
\qquad {}
 \non
\eea
Moreover, the theories (\ref{scalar}) and (\ref{tensor}) are dual to each other, because they are related 
to each other by the first-order action \cite{Siegel}
\bea
S_{\text{first-order}}=   -\int \rd^4 x\,{\rm d}^4\q \, 
\Big\{\,
\hf \P^2 
-\P( \J  + {\bar \J})  
\Big\}~.
\eea
Here $\P$ is a real unconstrained superfield.

As a generalization of the above example, consider
a model of $n$ self-interacting tensor multiplets \cite{Siegel}
\bea
S&=&  \int \rd^4 x\,{\rm d}^4\q \, \cL(G^i)~.
\label{Tensor-general}
\eea
It proves to be  dual to a nonlinear sigma-model described by chiral scalars $\J_i$ and their
conjugates ${\bar \J}_i$ \cite{LR}.  To  construct the latter, one introduces the first-order 
action \cite{LR} 
\bea
S_{\text{first-order}}=   \int \rd^4 x\,{\rm d}^4\q \, 
\Big\{\,
\cL (\P^i  )
+\P^i( \J_i  + {\bar \J}_i)  
\Big\}~, 
\eea
where $\P^i$ are real unconstrained superfields.
Varying $\J_i$ gives $\P^i = G^i$, and then $ S_{\text{first-order}}$ reduces
to (\ref{Tensor-general}). On the other hand, one can integrate out $\P$s using 
their equations of motion
\bea
\frac{\pa}{\pa \P^i} \cL(\P) + \J_i +{\bar \J}_i =0~,
\eea
to end up with 
\bea
S&=&  \int \rd^4 x\,{\rm d}^4\q \, \cK(\J^i +{\bar \J}^i)~, 
\qquad 
\cK(\J +{\bar \J}) := \cL (\P  )
+\P^i( \J_i  + {\bar \J}_i)  ~.~~~
\eea
The K\"ahler potential, $\cK$ is the Legendre transform of  $\cL$. Since $\cK$ depends on 
$\J$s and ${\bar \J}$s only via combinations $(\J +{\bar \J})$s, the $2n$-dimensional 
target spaces possesses  at least $n$ U(1) isometries.

The Legendre transformation considered generalizes to any number of tensor multiplets 
interacting with matter \cite{LR}. In particular, it can be applied to the model of our interest, 
eq. (\ref{TensorAction2}). This requires  considering  the following first-order action:
\bea
S_{\rm first-order}=   \int \rd^4 x\,{\rm d}^4\q \, 
\Big\{\,
L (\vf^i , {\bar \vf}^i, \P^i )
+\P^i( \J_i  + {\bar \J}_i)  
\Big\}~.
\eea
Here $\P^i$ is real unconstrained, and $\J_i$ is chiral, ${\bar D}_{\dt \a} \J_i=0$.
Integrating out the variables $\P$s  leads to an action of the form
\bea
S_{\rm dual}=   \int \rd^4 x\,{\rm d}^4\q \, 
H(\vf^i , {\bar \vf}^i, \J_j + {\bar \J}_j)~,
\label{7.38}
\eea
where $H(\vf , \bar \vf , \J + \bar \J)$  is the Legendre transform of
$L (\vf , \bar \vf , G )$.
It is the K\"ahler potential of a hyperk\"ahler manifold.

To construct a dual of (\ref{7.38}), with respect to the variables $\J$s and $\bar \J$s,
one could again use a first-order action of the type (\ref{f-o3}), 
that is
\bea
S_{\text{first-order}}=   \int \rd^4 x\,{\rm d}^4\q \, 
\Big\{ 
H(\vf^i , {\bar \vf}^i, U_j + {\bar U}_j)~
-\S^i  U_i - {\bar \S}^i {\bar U}_i 
\Big\}~,
\eea
where the superfields $U_i$ are complex unconstrained, and $\S^i$ complex linear.
However, the equations of motion for $U$s and $\bar U$s imply that $\S^i= {\bar \S}^i =G^i$.
Therefore, the duality transformation can be performed using the following first-order action:
\bea
S_{\text{first-order}}=   \int \rd^4 x\,{\rm d}^4\q \, 
\Big\{ 
H(\vf^i , {\bar \vf}^i, V_j )~
-G^i  V_i 
\Big\}~,
\eea
with the variables $V_i$ real unconstrained.

\subsection{Universality of polar multiplet sigma-models}
In general, off-shell $\cN=2$ supersymmetric $\s$-models can describe couplings 
of tensor multiplets, $\cO(2n)$ multiplets and polar multiplets.
\bea
S= \frac{1 }{ 2\pi\ri }
 \oint 
 \frac{\rd\z }{  \z}
\int\rd^4 x\,{\rm d}^4\q \,
\cL\Big(\eta (\z), H^{[\dots ]} (\z) , \U (\z) , \breve{\U} (\z) ; 
\z \Big)~.
\eea
However, it is always possible, in principle, to dualize any tensor multiplet into a polar multiplet, 
and also any  $\cO(2n)$ multiplet into a  polar one \cite{LR-projective1,G-RRWLvU}.
As a result, the most general $\cN=2$ $\s$-model can in principle be described  by polar multiplets only, 
using the action \cite{LR-projective1}
\bea
S= \frac{1 }{ 2\pi\ri }
 \oint 
 \frac{\rd\z }{  \z}
\int\rd^4 x\,{\rm d}^4\q \,
\cL\Big( \U (\z) , \breve{\U} (\z) ; 
\z \Big)~.
\label{PolarMostGen}
\eea
Different choices of $\cL\Big( \U, \breve{\U}; \z \Big)$ may lead to one and the same hyperk\"ahler 
geometry. The point is that a polar multiplet can be dualized into a polar one \cite{GK1,LR2008}, 
and the dual Lagrangian  differs, in general, from  the original one.

Example: For any real parameter $\a \in \mathbb R$,   $\a \neq \pm1$,
the Lagrangian 
\bea
\cL_\a \Big( \U, \breve{\U}; 
{\z} \Big)= \frac{1}{1-\a^2} \Big\{  \breve{\U} \U + \frac{ \a}{2} \Big( \frac{1}{\z^2} \U^2 
+ { \z^2} \breve{\U}^2\Big) \Big\}
\eea
is equivalent (dual) to the free polar multiplet Lagrangian
\bea
\cL \big( \U, \breve{\U}  \big)=   \breve{\U} \U~.
\eea

\section{$\cN = 2$ supersymmetric sigma-models on cotangent bundles of K\"ahler manifolds}
\setcounter{equation}{0}
As discussed earier, the most general sigma-model couplings of polar multiplets, 
eq. (\ref{PolarMostGen}), were introduced in 1988 by Lindstr\"om and Ro\v{c}ek \cite{LR-projective1}.
For some ten years, this theory remained   a purely formal construction, because  
there existed no technique to eliminate the auxiliary superfields contained in the arctic 
multiplet, except in the case of  Lagrangians quadratic in $\U $ and $\breve{\U}$.
This situation changed  in the late 1990s when  Refs. \cite{K98,GK1,GK2} 
identified a subclass of  models (\ref{PolarMostGen}),
possessing interesting geometric properties.
For these models one can develop a simple
procedure to eliminate the auxiliaries in perturbation theory, and in some cases exactly.
They are described by $\cN=2$ supersymmetric actions of the form:
\bea
S[\U, \breve{\U}]  =  
\frac{1}{2\pi {\rm i}} \, \oint_{ \g} \frac{{\rm d}\z}{\z} \,  
 \int   \rd^4 x\,{\rm d}^4\q \, 
K \Big( \U^I (\z), \breve{\U}^{\bar{J}} (\z)  \Big) ~,
\label{nact} 
\eea
where $\g$ denotes a closed contour around the origin, 
$K(\Phi^{I},{\bar \Phi}{}^{\bar{J}})$ is  the  K\"ahler potential of a real-analytic K\"ahler manifold $\cM$.

As usual, the dynamical variables $\U^I (\z)$ and $\breve{\U}^{\bar J} (\z)$  in (\ref{nact}) 
are arctic and antarctic multiplets, respectively:
\bea
 \U^I (\z) = \sum_{n=0}^{\infty}  \, \U^I_n \z^n = 
\F^I + \S^I \,\z+ O(\z^2) ~,\qquad
\breve{\U}^{\bar J} (\z) = \sum_{n=0}^{\infty}  \, {\bar
\U}^{\bar J}_n
 (-\z)^{-n}~.~~~
\label{exp}
\eea
Here $\F^I$ is chiral, $\S^I$  complex linear, 
\bea
{\bar D}_{\dt{\a}} \F^I =0~, \qquad  {\bar D}^2 \S^I = 0 ~,
\eea
and the remaining component superfields are  complex unconstrained.
The above theory 
is a minimal $\cN=2$ extension of the 
general  $\cN=1$ supersymmetric 
nonlinear sigma-model \cite{Zumino}
\bea
S[\F, \bar \F] =  \int 
\rd^4 x\,{\rm d}^4\q
\, K(\Phi^{I},
 {\bar \Phi}{}^{\bar{J}})  ~.
\label{nact4}
\eea

\subsection{Geometric properties}
Let us turn to discussing the geometric properties of  the theory (\ref{nact}).
What distinguishes   the Lagrangian in (\ref{nact}) from that appearing in 
the most general case, eq. (\ref{PolarMostGen}),  
is that the former has no explicit dependence on $\z$.
As is well-known from Classical Mechanics, the mathematical realization 
of the principle of the homogeneity of time is that the Lagrangian of a closed dynamical system  has no explicit dependence on  the time variable. Given such a Lagrangian, $L(q, {\dt q} )$, 
the action is invariant under arbitrary time translations. 
The Lagrangian in   (\ref{nact}) has no explicit dependence of $\z$ which can be viewed to be a complex 
evolution parameter. It is easy to see that (\ref{nact})  
is invariant under U(1) transformations 
\bea
\U(\zeta) ~~ \mapsto ~~ \U({\rm e}^{{\rm i} \a} \zeta) 
\quad \Longleftrightarrow \quad 
\U_n(z) ~~ \mapsto ~~ {\rm e}^{{\rm i} n \a} \U_n(z) ~.
\label{rfiber}
\eea
Transformations $ \z \to {\rm e}^{{\rm i} \a} \zeta$ can be interpreted as 
 time translations along $\g$. This becomes manifest if  the integration  contour 
 $\g$  is chosen to be  $\z(t)= R \,{\rm e}^{{\rm i} t}$.

The $\cN=2$  supersymmetric  sigma-model  
inherits  all the geometric features of its $\cN=1$ predecessor, specifically: 
\begin{itemize}
\item {\it K\"ahler invariance}
\begin{subequations}
\bea
\cN &=&1 ~ {\rm case}: \qquad
K(\F, \bar \F) \quad \longrightarrow \quad K(\F, \bar \F) +
\L(\F)+  {\bar \L} (\bar \F) ~,\\ 
\cN &=& 2  ~ {\rm case}: \qquad
K(\U, \breve{\U})  \quad \longrightarrow \quad K(\U, \breve{\U}) +
\L(\U) + {\bar \L} (\breve{\U} ) ~;~~~~
\label{kahl2}
\eea
\end{subequations}
\item
{\it Holomorphic reparametrizations of the K\"ahler manifold}
\begin{subequations}
\bea
\cN &=&1 ~ {\rm case}: \qquad
\F^I  \quad  \longrightarrow  \quad \F'{}^I=f^I \big( \F \big) ~, \\
\cN &=&2 ~ {\rm case}: \qquad
\U^I (\z) \quad  \longrightarrow  \quad \U'{}^I(\z)=f^I \Big (\U(\z) \Big)~.
\label{kahl3}
\eea
\end{subequations}
\end{itemize}
Therefore, the physical
superfields of the 
${\cal N}=2$ theory
\bea
 \U^I (\z)\Big|_{\z=0} ~=~ \F^I ~,\qquad  \quad \frac{ {\rm d} \U^I (\z) 
}{ {\rm d} \z} \Big|_{\z=0} ~=~ \S^I ~,
\label{kahl4} 
\eea
should be regarded, respectively, as  coordinates of a point in the K\" ahler
manifold and a tangent vector at  the same point. We conclude that
the variables $(\F^I, \S^J)$ parametrize the holomorphic tangent 
bundle $T\cM$ of the K\"ahler manifold \cite{K98}.

\subsection{Tangent-bundle and cotangent-bundle formulations}
To describe the theory in terms of 
the physical superfields $\F^I$ and $\S^I$ only, 
all the auxiliary superfields have to be eliminated  with the aid of the 
corresponding algebraic equations of motion
\bea
\oint \frac{{\rm d} \z}{\z} \,\z^n \, \frac{\pa K(\U, \widetilde{\U} ) }{\pa \U^I} 
~ = ~ \oint \frac{{\rm d} \z}{\z} \,\z^{-n} \, \frac{\pa 
K(\U, \widetilde{\U} ) } {\pa \widetilde{\U}^{\bar J} } 
~ = ~ 0 ~, \qquad n \geq 2 ~ .               
\label{asfem}
\eea
Let $ \U^I_*(\z) \equiv \U_*( \z; \F, {\bar \F}, \S, \bar \S )$ 
denote their  {unique solution subject to the initial conditions}
\bea
\U^I_* (0)  = \F^I ~,\qquad  \quad \dt{\U}{}_*^I (0) 
 = \S^I ~.
\label{geo3} 
\eea

${}$For a general K\"ahler manifold $\cM$, 
the auxiliary superfields $\U^I_2, \U^I_3, \dots$, and their
conjugates,  can be eliminated using perturbation theory only,  
with the aid 
of  the following {\it ansatz} \cite{KL}:
\bea
\U^I_n = 
\sum_{p=0}^{\infty} 
G^I{}_{J_1 \dots J_{n+p} \, \bar{L}_1 \dots  \bar{L}_p} (\F, {\bar \F})\,
\S^{J_1} \dots \S^{J_{n+p}} \,
{\bar \S}^{ {\bar L}_1 } \dots {\bar \S}^{ {\bar L}_p }~, 
\qquad n\geq 2~.~~~~~
\label{ansatz}
\eea
This is the most general ansatz compatible with the U(1) symmetry 
(\ref{rfiber}). The only essential assumption to justify the use of perturbation theory is
the requirement that the K\"ahler potential $K(\F, \bar \F)$ is real analytic.
Determining step by step the coefficients $G^I{}_{J_1 \dots J_{n+p} \, \bar{L}_1 \dots  \bar{L}_p} (\F, {\bar \F})$,
we can completely reconstruct the required solution $ \U^I_*(\z) $.

In some cases, the  solution $ \U^I_*(\z) $ can be determined exactly. 
Let $\cM$ be  a Hermitian symmetric space, and hence its curvature tensor is covariantly constant.
\bea
\nabla_L  R_{I_1 {\bar  J}_1 I_2 {\bar J}_2}
= {\bar \nabla}_{\bar L} R_{I_1 {\bar  J}_1 I_2 {\bar J}_2} =0~.
\label{cc}
\eea
Then, the curve $\U_*(\z)$ turns out to obey the  generalized geodesic equation \cite{GK1}:
\bea
\frac{ {\rm d}^2 \U^I_* (\z) }{ {\rm d} \z^2 } + 
\G^I_{JK} \Big( \U_* (\z), \bar{\F} \Big)\,
\frac{ {\rm d} \U^J_* (\z) }{ {\rm d} \z } \,
\frac{ {\rm d} \U^K_* (\z) }{ {\rm d} \z } =0~.
\label{8.13}
\eea
A derivation of this result will be given below. 
It follows from (\ref{8.13}) that only the term with $p=0$ in (\ref{ansatz}) is non-zero
in the case of Hermitian symmetric spaces.

Suppose that  all the auxiliary superfields 
have been eliminated, and the $\U_*(\z)$ is known explicitly. 
The next technical problem  to address is the  evaluation of  the contour  integral:
\bea 
S_{{\rm tb}}[\F,  \S]  :=  
\frac{1}{2\pi {\rm i}} \, \oint \frac{{\rm d}\z}{\z} \,  
 \int \rd^4 x\,{\rm d}^4\q \, 
K \big( \U_* (\z), \breve{\U}_* (\z)  \big) ~.
\eea
This is  only a technical issue, rather complicated in practice.  
However complicated, the outcome should be an action of the form:
\bea
S_{{\rm tb}}[\F,  \S] 
&=& \int 
\rd^4 x\,{\rm d}^4\q
\, \Big\{\,{
K \big( \F, \bar{\F} \big)+  
\cL \big(\F, \bar \F, \S , \bar \S \big)}\Big\}~,\non \\
\cL 
&=&
\sum_{n=1}^{\infty}  \cL_{I_1 \cdots I_n {\bar J}_1 \cdots {\bar 
J}_n }  \big( \F, \bar{\F} \big) \S^{I_1} \dots \S^{I_n} 
{\bar \S}^{ {\bar J}_1 } \dots {\bar \S}^{ {\bar J}_n }
:= \sum_{n=1}^{\infty}  \cL^{(n)} ~.~~~~~~~~
\label{act-tab-mod}
\eea
Here $\cL_{I {\bar J} }=   -g_{I \bar{J}} \big( \F, \bar{\F}  \big) $ 
and the  coefficients $\cL_{I_1 \cdots I_n {\bar J}_1 \cdots {\bar 
J}_n }$, for  $n>1$, 
are tensor functions of the K\"ahler metric
$g_{I \bar{J}} \big( \F, \bar{\F}  \big) 
= \pa_I 
\pa_ {\bar J}K ( \F , \bar{\F} )$, 
 the Riemann curvature $R_{I {\bar 
J} K {\bar L}} \big( \F, \bar{\F} \big) $ and its covariant 
derivatives.  
Each term $ \cL^{(n)} $ in the Lagrangian contains equal powers
of $\S$ and $\bar \S$, since the original action
is invariant under the rigid U(1)  transformations (\ref{rfiber}).

It is instructive to reproduce here  the explicit expressions for several functions $\cL^{(n)}$
appearing in (\ref{act-tab-mod}). Direct calculations give
\begin{subequations}
\bea
\cL^{(1)} &=&   -g_{I \bar{J}} \S^I {\bar \S}^{\bar J}~, 
\label{L1}  \\
\cL^{(2)} &=& \frac{1}{4} R_{I_1 {\bar J}_1 I_2 {\bar J}_2} \S^{I_1}\S^{I_2}
{\bar \S}^{{\bar J}_1}{\bar \S}^{{\bar J}_2}~,
\label{L2}   \\
\cL^{(3)} &=&- \frac{1}{12} \Big\{ \frac{1}{6} 
\{ \nabla_{I_3}, {\bar \nabla}_{{\bar J}_3} \}
R_{I_1 {\bar J}_1 I_2 {\bar J}_2} 
+R_{I_1 {\bar J}_1 I_2 }{}^LR_{L {\bar J}_2 I_3 {\bar J}_3}\Big\}  \non \\
&&\qquad  \qquad \times
\S^{I_1}\dots \S^{I_3}
{\bar \S}^{{\bar J}_1}\dots{\bar \S}^{{\bar J}_3}
\label{L3}
\eea
\end{subequations}
The expression for $\cL^{(4)}$ is given in \cite{K-hyper}.

To construct the dual formulation of (\ref{act-tab-mod}), 
we follow the general scheme  of subsection \ref{GeneralizedLT}
and consider the first-order action
\bea
S_{\text{first-order}}=   \int \rd^4 x\,{\rm d}^4\q \, 
\Big\{\,
K \big( \F, \bar{\F} \big)+  \cL
\big(\F, \bar \F, \G , \bar \G \big)
+\J_I \,\G^I + {\bar \J}_{\bar I} {\bar \G}^{\bar I} 
\Big\}~.~~
\label{f-o}
\eea
Here the tangent vector $\G^I$ is now  complex unconstrained, 
while the one-form  
$\Psi_I$ is chiral, 
${\bar D}_{\dt \a} \J_I =0$.
Varying $\J_I$ gives ${\bar D}^2 \G^I =0$, that is $\G^I =\S^I$,
and then (\ref{f-o}) reduces to the original action.
On the other hand, varying $\G^I$ gives
\bea
\frac{\pa }{\pa \G^I} \cL \big(\F, \bar \F, \G , \bar \G \big)+\J_I =0~.
\label{f-o-equation}
\eea
Eliminating the $\G$s and their conjugates\footnote{Since 
$\cL=   -g_{I \bar{J}} \S^I {\bar \S}^{\bar J} +O(|\S|^4)$, 
both requirements (\ref{Hessian1}) and (\ref{Hessian4}) hold 
in a neighborhood of the zero section of the tangent bundle,  $T\cM$, of $\cM$.}
 leads to the dual action
\bea
S_{{\rm ctb}}[\F,  \J]  
&=&   \int \rd^4 x\,{\rm d}^4\q \, 
\Big\{
K \big( \F, \bar{\F} \big)+    
\cH \big(\F, \bar \F, \J , \bar \J \big)\Big\}~,\non\\
  \cH \big(\F, \bar \F, \J , \bar \J \big)&=&
\sum_{n=1}^{\infty} \cH^{I_1 \cdots I_n {\bar J}_1 \cdots {\bar 
J}_n }  \big( \F, \bar{\F} \big) \J_{I_1} \dots \J_{I_n} 
{\bar \J}_{ {\bar J}_1 } \dots {\bar \J}_{ {\bar J}_n } ~,
\label{act-ctb}
\eea
with
\bea
&&\cH^{I {\bar J}} \big( \F, \bar{\F} \big) 
= g^{I {\bar J}} \big( \F, \bar{\F} \big) ~.
\eea
The fact that each term in the expansion of $ \cH \big(\F, \bar \F, \J , \bar \J \big)$ contains equal powers
of $\J$ and $\bar \J$, 
follows from the invariance of (\ref{f-o}) 
 under the rigid U(1)  transformations
\bea
\F^I(z) ~ \to ~    \F^I(z) ~, 
\quad
\G^I(z) ~ \to ~ {\rm e}^{{\rm i}  \a} \G^I(z) ~, 
\quad \J_I(z) ~ \to ~ {\rm e}^{-{\rm i}  \a} \J_II(z)~. 
\eea

In the dual formulation of the $\cN=2$ supersymmetric sigma-model, 
the target space  is (an open neighborhood  of the zero section of)
 the cotangent bundle $T^*\cM$ of the K\"ahler manifold $\cM$ \cite{GK1,GK2}.
It is therefore a hyperk\"ahler space, and 
\bea
 {\mathbb K} (\F, \bar \F, \J, \bar \J )
:=K \big( \F, \bar{\F} \big)+  \cH(\F, \bar \F, \J, \bar \J )
\eea
the corresponding hyperk\"ahler  potential. Since 
\bea
\cH(\F, \bar \F, \J, \bar \J )
= g^{I {\bar J}} \big( \F, \bar{\F} \big) \J_I {\bar \J}_{\bar J} +O(|\J|^4)~, 
\eea
the hyperk\"ahler metric is nonsingular in  a neighborhood  of the zero section of $T^*\cM$.
These results agree with those derived independently in the mathematical 
literature by purely geometric means \cite{Kaledin,Feix}.

\subsection{Hermitian symmetric spaces: Method I}
If the K\"ahler manifold $\cM$ is Hermitian symmetric, 
then the $\cN=2$ supersymmetric sigma-model on $T^*\cM$ 
can be derived in closed form, as was first sketched in \cite{GK1}.
To carry out such a construction, 
there have been developed two alternative methods that are based on the use of conceptually 
different ideas and tools:

\begin{itemize}

\item Method I  \cite{GK1,GK2,AN,AKL1}
makes use of the properties  that 
(i) $\cM$ is a homogeneous space, $\cM = G/H$;  
(ii) the group $G$ acts on $\cM$ by holomorphic isometries.

\item {Method II}  \cite{AKL2,KN} makes use of
(i) the covariant constancy of the curvature;
(ii) extended supersymmetry.
\end{itemize}
We now turn to discussing the first method. Method II will be reviewed in the next subsection.

As before, denote by $\U_*(\z) \equiv \U_*( \z; { \F, {\bar \F}, \S, \bar \S} )$ 
the unique solution of the auxiliary field equations (\ref{asfem})
under  the initial conditions (\ref{geo3}).
Using the canonical coordinates \cite{Bochner,Calabi}
for the Hermitian symmetric space $\cM$, which are defined in Appendix B,
 we can find a part of the solution:
\bea
 \U_0(\z) \equiv \U_{p_0} (\z)
:=   \U_*( \z; { \F =0, {\bar \F} =0 , \S_0, {\bar \S}_0} )~, \qquad \U_{p_0} (0)=p_0
\eea
with  $\S_0$ a tangent vector at $p_0 \in \cM$, the origin of the canonical coordinate system. 
It is 
\bea
\U_0 (\z) = \S_0 \, \z~, \qquad  
\breve{\U}_0 (\z) =-  \frac{\bar{\S}_0}{\z}~.
\label{u-0}
\eea

As a next step, we can construct a curve 
$$\U_p (\z)~, \qquad \U_p (0) =p \in \cM
$$obtained from $\U_{p_0}(\z)$ by applying an isometry transformation 
$g \in G$ such that $g\cdot p_0 =p$.
{The holomorphic isometry transformations 
leave invariant the auxiliary field equations.} 

Let $U\subset \cM$ be the neighborhood on which 
the canonical coordinate system is defined.
We can construct a coset representative, $\cS$:  $U \to  G$, with the following property: 
associated with a point $p \in U$ is 
the holomorphic isometry  $\cS[p] \in G$ of $\cM$,
${}q  \to \cS[p]\cdot q \in \cM$,  for any $ q \in \cM$,
such that 
\be
\cS[p] \cdot p_0 =p~. 
\ee
In local coordinates, $\cS[p] = \cS[\F, {\bar \F}]$, and it acts on a generic point 
 $q \in U$ parametrized by complex variables $(\X^I , {\bar \X}^{\bar J} )$ as follows:
\bea
\X~ \to~ \X' = f (\X; \F, \bar \F )~, \qquad 
f (0; \F, \bar \F ) =\F~.
\eea
Now, applying the group transformation $\cS(\F, \bar \F )$ 
to $\U_{0}(\z)$  gives 
\bea 
\U_0(\z) ~\to ~ \U_*(\z) =  f (\U_0(\z) ; \F, \bar \F )
= f (\S_0 \,\z ; \F, \bar \F )~, \qquad 
 \U_*(0)=\F~. 
 \eea
Imposing the second initial condition, $ \dt{\U}_* (0) = \S $, gives
\bea
\S^I =\S^J_0 \, \frac{\pa }{\pa \X^J}
f^I (\X; \F, \bar \F ) \Big|_{\X=0} ~,
\label{generalizedlinear}
\eea
and thus $\S_0$ can be uniquely expressed in terms of $\S$ and $\F$, $\bar \F$.
As a result, the desired curve by $\U_*(\z) $ has been constructed.

As pointed out earlier, the curve  $\U_*(\z) $ obeys the  generalized geodesic equation (\ref{8.13}).
Now we  are in a position to justify the claim.
In the canonical coordinate system, the curve
\bea
\U_0 (\z) = \S_0 \, \z~, \qquad  
\breve{\U}_0 (\z) =-  \frac{\bar{\S}_0}{\z}
\eea
is characterized by
\bea
\frac{ {\rm d}^2 \U^I_0 (\z) }{ {\rm d} \z^2 } =
\frac{ {\rm d}^2 \U^I_0 (\z) }{ {\rm d} \z^2 } + 
\G^I_{JK} \Big( \U_0 (\z), \bar{\F}=0 \Big)\,
\frac{ {\rm d} \U^J_0 (\z) }{ {\rm d} \z } \,
\frac{ {\rm d} \U^K_0 (\z) }{ {\rm d} \z } =0~. 
\eea
Since the equation
\bea
\frac{ {\rm d}^2 \U^I (\z) }{ {\rm d} \z^2 } + 
\G^I_{JK} \Big( \U (\z), \bar{\F} \Big)\,
\frac{ {\rm d} \U^J (\z) }{ {\rm d} \z } \,
\frac{ {\rm d} \U^K (\z) }{ {\rm d} \z } =0~ 
\eea
is invariant under holomorphic isometries, we conclude that 
 $\U_*(\z)$ indeed obeys the  generalized geodesic equation (\ref{8.13}).

Eq. (\ref{8.13}) leads to a simple corollary that is  of special importance for 
the method to be discussed in the next subsection. Using repeatedly eq. (\ref{8.13}) allows us to compute 
the Taylor coefficients $\U_2, \U_3, \dots, $ in the expansion
\bea
\U^I_* (\z) = \sum_{n=0}^{\infty}  \, \U^I_n \z^n = 
\F^I + \S^I \,\z+ \U^I_2 \, \z^2 +O(\z^3) ~.
\eea
In particular, we derive
\bea
\U^I_2 = -\hf \G^I_{~JK} \big( \F, \bar{\F} \big) \, \S^J\S^K~.
\label{U2}
\eea

\subsection{The Eguchi-Hanson metric and its non-compact cousin}

As an instructive application of the method described, we 
consider the two-sphere $\cM = {\rm SU(2)/U(1) } \cong
{\mathbb C}P^1 ={\mathbb C} \, \cup
\,\{\infty \}$.
In the north chart of ${\mathbb C}P^1 $, the K\"ahler potential and 
metric are:
\bea
K (z, {\bar z}) = r^2 \ln \left(1 + {z {\bar z}} \right) ~, \qquad
g_{z {\bar z}} (z, \bar z) = r^2  \left(1 + {z {\bar z} }  
\right)^{-2} ~,
\eea
with $1/r^2$ being proportional to the curvature.  
The coordinate system under consideration is canonical in the sense of Appendix B.

Fractional linear (isometry) transformation
\bea
z ~\longrightarrow ~\cS_{[\F,\bar \F]} (z) ~=~  \frac{z + \F   }
{- {\bar \F} z   +1}~, \qquad \cS_{[\F, \bar \F]} (0) ~=~ \F
\eea
induces
\bea
\U_* (\z) ~=~ \frac{\F (1 + \F {\bar \F}) + \z \S  }
{1 + \F {\bar \F} - \z {\bar \F} \S  }~,
\label{s2soln}
\eea
and then
\bea
K \left( \U_* (\z), \breve{\U}_* (\z) \right) \,&=& \, r^2 \ln 
\left\{ \Big( 1 + \F {\bar \F} \Big) \Big( 1 -  \frac{ 
\S \,{\bar \S} } { ( 1 + \F {\bar \F} )^2 }  \Big) \right\} \non \\
&&+ \z\, \L \big( \z \big) -\frac{1}{\z}\, {\bar \L} \big( -1 / \z  \big)~,
\eea
with $\L(\z)$ some holomorphic function.
The action becomes
\bea
S[\U_*, \breve{\U}_*] = \int {\rm d}^4x\,
{\rm d}^4 \q \, \left\{ 
K(\F, \bar 
\F) +  r^2 \ln \Big(1 - \frac{1}{r^2} \, g_{\F {\bar \F}} (\F, {\bar \F})\;
\S \,\bar \S \Big) \right\} ~,
\label{physpol}
\eea
and is well-defined under the global restriction
\bea
g_{\F {\bar \F}} (\F, {\bar \F})\, \S \, \bar \S  ~<~r^2~.
\label{domain}
\eea

To construct the dual formulation of (\ref{physpol}), we should introduce
the corresponding first-order action (\ref{f-o}). 
The equation of motion for $\G$, (\ref{f-o-equation}),  in our case becomes
\bea
\frac{g_{\F \bar{\F}}(\F, {\bar \F})\, \bar \G }
{r^2 - g_{\F \bar{\F}}(\F, {\bar \F}) \,\G \bar \G }
=\frac{1}{r^2} \,\J ~, \qquad 
g_{\F {\bar \F}} (\F, {\bar \F})\, \G\, \bar \G  ~<~r^2~.
\label{ball1}
\eea
This equation and its conjugate allow us to express $\G$ in terms of $\J$ and its 
conjugate, without any restriction on $\J$\footnote{The relations (\ref{domain}) and 
(\ref{mom}) are analogous to those appearing 
in special relativity. For a massive particle,
its velocity, $\vec{v}$, is constrained by $| \vec{v}| <c$, however the three-momentum, 
$\vec{p}$, can take arbitrary values, $| \vec{p}| <\infty$.}
\bea
 \J\, \bar \J  ~<~\infty~.
\label{mom}
\eea
As a result, the target space of the dual theory coincides with  $T^*{\mathbb C}P^1$
parametrized by local complex variables $(\F, \J)$.
The Lagrangian of the dual theory, eq. (\ref{act-ctb}), is 
the hyperk\"ahler potential
\begin{subequations}
\bea
{\mathbb K} ( \F, \bar{\F} , \J , \bar \J ) &=&
 K(\F, \bar \F) +  \cH (\F, \bar \F, \J, \bar \J)~,
 \non \\
 \cH (\F,\bar \F, \J, \bar \J) &=& 
r^2 \k \,\cF (\k )~, \qquad 
r^2  \k =
g^{\F {\bar \F}} (\F, {\bar \F}) \,\J \bar \J ~, \\
 \cF(x) &:=& 
 \frac{1}{x} \,\Big\{ \sqrt{1+4x} -1 -\ln \frac{1+ \sqrt{1+4x} }{2} \Big\}~, 
\qquad \cF(0)=1~~~~~~~
\label{F-Calabi}
\eea
\end{subequations}
corresponding to the Eguchi-Hanson metric in the form given by Calabi \cite{Calabi2}.

It is of some interest to repeat the above analysis for the 
complex hyperbolic line
$\cM = {\rm SU(1,1) /U (1)} \equiv \bH$, which is a non-compact cousin
of ${\mathbb C}P^1 $.
The K\"ahler potential and  metric of $\bH$ are: 
\bea
K (z, {\bar z}) ~=~ -r^2 \ln \left(1 - {z {\bar z}} \right) ~, \qquad
g_{z {\bar z}} (z, \bar z) ~=~ r^2\, \left(1 - {z {\bar z} }  
\right)^{-2} ~.
\eea
Instead of the action (\ref{physpol}), we now have 
\bea
S[\U_*, \breve{\U}_*] &=& \int {\rm d}^4x\,
{\rm d}^4 \q \, \left\{ 
K(\F, \bar 
\F) -  r^2 \ln \Big(1 + \frac{1}{r^2} \, g_{\F {\bar \F}} (\F, {\bar \F})\;
\S \,\bar \S \Big) \right\} ~.~~~
\eea
The action is defined on $T \bH$, and {\it no restriction} on the tangent variable $\S$ occurs. 
However, the dual formulation of the theory  is well defined under the restriction
\bea
g^{\F {\bar \F}} (\F, {\bar \F})\, \J \, \bar \J  ~<~r^2~.
\eea
As a result, the hyperk\"ahler structure is defined on the open disc bundle
in the cotangent bundle $T^* \bH$.

It is known that any compact Riemann surface,  $\cM_g$, of genus $g>1$ can be obtained 
from $\bH$ by factorization  with respect to a discrete 
subgroup of ${\rm SU}(1,1)$, see, e.g., \cite{FK}. 
Using the hyperk\"ahler metric constructed on the open disc bundle in 
 $T^*\bH$, we  can generate a hyperk\"ahler 
 structure defined on  an open neighbourhood of the zero section 
 of $T^* \cM_g$. The obtained  hyperk\"ahler metric is not complete.

\subsection{Hermitian symmetric spaces: Method II}
{Method I} was successfully applied to 
the four series of compact Hermitian symmetric spaces:
\bea 
\frac{ {\rm U}(m+n) }{ {\rm U}(m) \times {\rm U}(n)}  ~, & \qquad  &
\frac{ { \rm Sp}(n)}{   {\rm U}(n)} ~, \qquad \frac{ {\rm SO}(2n) }{ {\rm U}(n) }~,
\qquad \frac{ {\rm SO} (n + 2) }{ {\rm SO}(n) \times {\rm SO}(2)}~,
\eea  
as well as to their non-compact versions:
\bea
\frac{ {\rm U}(m,n) }{ {\rm U}(m) \times {\rm U}(n) } ~, \qquad 
 \frac{ {\rm Sp}(n, {\mathbb R}) } {{\rm U}(n)  } ~, \qquad 
 \frac{ {\rm SO}^*(2n) }{{\rm U}(n) } ~, \qquad
\frac{ {\rm SO}_0(n , 2) }{ {\rm SO}(n) \times {\rm SO}(2)} ~,\eea
on case by case basis. This construction was finalized in \cite{AKL1}.
The general results are as follows:
\begin{itemize}
\item If the Hermitian symmetric space $\cM$ is compact, then the hyperk\"ahler structure is defined 
on the whole $T^*\cM$;

\item If  $\cM$ is non-compact, then the hyperk\"ahler structure is defined 
on a neighbourhood of the zero section of $T^*\cM$, and cannot be extended to the whole
cotangent bundle.
\end{itemize}
A detailed discussion of these are related properties can be found in \cite{AKL1}.

Although  {method I} worked well for the regular series listed,  it turned out 
to be impractical in the case of the exceptional Hermitian symmetric spaces
\bea
\frac{\rm E_6}{\rm SO(10) \times U(1)}~, \qquad \frac{\rm E_7}{\rm E_6 \times U(1)}~.
\eea
In order to work out these cases, an alternative method was developed in \cite{AKL2}.

The method introduced  in \cite{AKL2} is based on the use of extended supersymmetry.
Let us start from  the $\cN=2$ supersymmetry transformation of the arctic multiplet
\bea
\d \U^I (\z)= {\rm i} \left(\ve^\a_i Q^i_\a +
{\bar \ve}^i_\ad {\bar Q}^\ad_i \right)  \U^I(\z)
\label{SUSY1}
\eea
considered as a $\cN=2$ superfield.
As a next step, we
reduce this transformation to $\cN=1$ superspace. Then, 
the second hidden supersymmetry proves  to act on the 
physical superfields $\F$ and $\S$ as
\bea
\d \F^I = {\bar \ve}_{\dt{\a}} {\bar D}^{\dt{\a}} \S^I~, \qquad 
\d \S^I = -\ve^\a D_\a \F^I +   {\bar \ve}_{\dt{\a}} {\bar D}^{\dt{\a}} \U^I_2~.
\eea 
Upon elimination of the auxiliary superfields, the component  $\U^I_2$ becomes
a function of the physical superfields, 
\bea
\U^I_2 = -\hf \G^I_{~JK} \big( \F, \bar{\F} \big) \, \S^J\S^K~.
\eea

The tangent-bundle action 
\bea
S_{{\rm tb}}[\F,  \S] 
&=& \int 
\rd^4 x\,{\rm d}^4\q
\, \Big\{
K \big( \F, \bar{\F} \big)+  
\cL \big(\F, \bar \F, \S , \bar \S \big)\Big\}~,\non \\
\cL 
&=&
\sum_{n=1}^{\infty}  \cL_{I_1 \cdots I_n {\bar J}_1 \cdots {\bar 
J}_n }  \big( \F, \bar{\F} \big) \S^{I_1} \dots \S^{I_n} 
{\bar \S}^{ {\bar J}_1 } \dots {\bar \S}^{ {\bar J}_n }
\label{act-tab}
\eea
has to be invariant under the above supersymmetry transformation. 
This is a highly nontrivial requirement.
Indeed, by making use of the fact the the Riemann curvature is covariantly constant,
\bea 
\nabla_L  R_{I_1 {\bar  J}_1 I_2 {\bar J}_2}
= {\bar \nabla}_{\bar L} R_{I_1 {\bar  J}_1 I_2 {\bar J}_2} =0~,
\eea
and hence 
\bea
\nabla_L \cL_{I_1 \cdots I_n {\bar J}_1 \cdots {\bar 
J}_n } =  {\bar \nabla}_{\bar L} 
\cL_{I_1 \cdots I_n {\bar J}_1 \cdots {\bar 
J}_n } =0~, 
\eea
we are able to show that the action is indeed supersymmetric provided 
the Lagrangian $\cL$ obeys the following
{\it linear}  differential equation: 
\bea
\hf \S^K \S^L\,R_{K {\bar J} L }{}^I\, \cL_I 
+  \cL_{\bar J}  +g_{I \bar{J}}\, \S^I =0~,
\qquad \cL_I :=  \frac{\pa }{\pa \S^I} \cL ~.
\label{MasterEq1}
\eea
A general solution of (\ref{MasterEq1}), which is compatible 
with the series expansion (\ref{act-tab}), 
was found in \cite{AKL2}.  It is
\bea
\cL \big(\F, \bar \F, \S , \bar \S \big)
&=& - g_{I \bar{J}} 
 {\bar \S}^{\bar{J}} \,
\frac{ {\rm e}^{\cR_{\S,{\bar \S}}} -1}{ \cR_{\S,{\bar \S} }}\,
 \S^I   ~, \non \\
 \cR_{\S,{\bar \S}} &:=&  -\hf \S^K {\bar \S}^{ {\bar L} } \,
R_{K {\bar L} I }{}^J   \,\S^I \frac{\pa}{\pa \S^J}~.~~~~
\label{closed}
\eea

In the dual, cotangent bundle formulation
\bea
S_{{\rm ctb}}[\F,  \J]  
&=& \int 
{\rm d}^4 x \,{\rm d}^4\q
\, \Big\{\,
K \big( \F, \bar{\F} \big)+    
\cH \big(\F, \bar \F, \J , \bar \J \big)\Big\}~,\non \\
\cH
&=& \sum_{n=1}^{\infty} 
\cH^{I_1 \cdots I_n {\bar J}_1 \cdots {\bar 
J}_n }  \big( \F, \bar{\F} \big) \J_{I_1} \dots \J_{I_n} 
{\bar \J}_{ {\bar J}_1 } \dots {\bar \J}_{ {\bar J}_n } ~,
\eea
the `Hamiltonian' $\cH$ proves to obey the {\it nonlinear} differential equation:
\bea
\cH^I \,  g_{I {\bar J}} - \hf \, \cH^K\cH^L \,  R_{K {\bar J} L}{}^I \,\J_I =
{\bar \J}_{ \bar J} ~, 
\qquad \cH^I  = \frac{\pa}{\pa \J_I}  \cH~.
\eea
It can be deduced from (\ref{MasterEq1})
by making use of the properties of the Legendre transformation.

Using the above results, the case of $E_6/ SO(10) \times U(1)$ was worked out 
for the first time in \cite{AKL2}.

A closed-form expression for $\cH(\F, \bar \F, \J , \bar \J ) $ was not obtained in \cite{AKL2}.
It was derived in  \cite{KN}. The same work also provided an alternative 
closed-form expression for $\cL \big( \F, \bar \F,  \S,  {\bar \S} \big) $.
We reproduce here only the results obtained in \cite{KN}.

\underline{Tangent-bundle formulation}
\bea
\cL \big( \F, \bar \F,  \S,  {\bar \S} \big) = - \hf {\bm \S}^{\rm T} {\bm g} \,
\frac{ \ln \big( {\mathbbm 1} + {\bm R}_{\S,\bar \S}\big)}{\bm R_{\S,\bar \S}}
\, {\bm \S}~, \qquad 
{\bm \S} :=\left(
\begin{array}{c}
\S^I\\
{\bar \S}^{\bar I} 
\end{array}
\right) ~.
\label{closed2}
\eea
Here we have defined
\bea
{\bm g} &:=&\left(
\begin{array}{cc}
0 & g_{I \bar J}\\
g_{{\bar I}J} &0 
\end{array}
\right)~, \qquad
{\bm R}_{\S,\bar \S}
:=\left(
\begin{array}{cc}
0 & (R_\S)^I{}_{\bar J}\\
(R_{\bar \S})^{\bar I}{}_J &0 
\end{array}
\right)~,  
\non \\
 (R_\S)^I{}_{\bar J}&:=&\hf R_K{}^I{}_{L \bar J}\, \S^K \S^L~, \qquad
 (R_{\bar \S})^{\bar I}{}_J := \overline{(R_\S)^I{}_{\bar J}}~.~~~~
\label{R-Sigma}
\eea

\underline{Cotangent-bundle formulation}
\bea
\cH(\F, \bar \F, \J , \bar \J ) = \hf {\bm \J}^{\rm T}{\bm g}^{-1}  \cF \Big( - {\bm R}_{\J,\bar \J} \Big)\, 
{\bm \J} ~, \qquad 
{\bm \J} :=\left(
\begin{array}{c}
\J_I\\
{\bar \J}_{\bar I} 
\end{array}
\right) ~,
\label{hyperkahler-potential}
\eea
where the function $\cF(x)$ is given by eq. (\ref{F-Calabi}).
The operator $ {\bm R}_{\J,\bar \J} $ is defined as
\bea
{\bm R}_{\J,\bar \J}
&:=&\left(
\begin{array}{cc}
0 & (R_\J)_I{}^{\bar J}\\
(R_{\bar \J})_{\bar I}{}^J &0 
\end{array}
\right)~, \non \\ 
(R_\J)_I{}^{\bar J}&=& (R_\J)_{IK} \,g^{K \bar J}~, \qquad 
(R_\J )_{K L}:= \hf R_{K}{}^I{}_{L}{}^J \,\J_I \J_J~. 
\eea

As a result, for any Hermitian symmetric space $\cM$,
the hyperk\"ahler potential on $T^*\cM$ is:
\bea
{\mathbb K} (\F, \bar \F, \J, \bar \J ) =
K(\F, \bar \F) +  \hf {\bm \J}^{\rm T}{\bm g}^{-1}  \cF \Big( - {\bm R}_{\J,\bar \J} \Big)\, 
{\bm \J} ~.
\eea

In the mathematical literature, there exists a different representation
for  the hyperk\"ahler potential  derived in \cite{BG}.
The  Biquard-Gauduchon representation is
\bea
\cH(\F, \bar \F, \J , \bar \J ) =  { \J}^\dagger \check{g}^{-1}  \cF \Big( - {\mathbb R}_{\J,\bar \J} \Big)\, 
{ \J} ~,  
\eea
where 
\bea
({\mathbb R}_{\J, \bar \J})_{I}{}^J := \hf R_I{}^{J \bar K L} \J_L {\bar \J}_{\bar K}~
\eea
and $ \check{g}$ denotes an off-diagonal block of the K\"ahler metric,
\bea
{\bm g}
:=\left(
\begin{array}{cc}
0 & g_{I \bar J}\\
g_{{\bar I}J} &0 
\end{array}
\right)
\equiv \left(
\begin{array}{cc}
0 & \hat{g}\\
\check{g} &0 
\end{array}
\right)~. 
\eea

The above unified formula was derived by Biquard and  Gauduchon
with the aid of  purely algebraic means involving
the root theory for Hermitian symmetric spaces,
in conjunction with some guesswork based on the use of the Calabi metrics 
for $T^* {\mathbb C}P^n$ \cite{Calabi2}.
In the supersymmetric setting described above, the results were obtained by making use of a regular procedure. 
No guesswork was needed.

\section{The case of an arbitrary K\"ahler manifold $\cM$} 
\setcounter{equation}{0}
In the previous section, we used the power of supersymmetry to determine 
the hyperk\"ahler potential on the cotangent bundle $T^*\cM$ of an arbitrary 
Hermitian symmetric space. It is natural to wonder how much information can be extracted 
by using similar supersymmetry considerations in the case when $\cM$ is an arbitrary 
real-analytic K\"ahler space. This problem was analyzed in \cite{K09}.
Here we give a brief review of the results obtained.

Upon elimination of the auxiliary superfields,
the second {hidden} supersymmetry becomes
\bea
\d \F^I = {\bar \ve}_{\dt{\a}} {\bar D}^{\dt{\a}} \S^I~, \qquad 
\d \S^I = -\ve^\a D_\a \F^I +   {\bar \ve}_{\dt{\a}} {\bar D}^{\dt{\a}} \,
{\U^I_2 (\F, \bar \F , \S , \bar \S)} ~.
\eea 
where the general form for ${\U^I_2 (\F, \bar \F , \S , \bar \S)} $ is as follows:
\bea
\U^I_2 (\F,  \bar \F, \S , \bar \S)&=& -\hf \G^I_{JK} \big( \F, \bar{\F} \big) \, \S^J\S^K+
 G^I(\F,  \bar \F, \S , \bar \S) ~, \non \\
 G^I(\F,  \bar \F, \S , \bar \S)&:=&
\sum_{p=1}^{\infty} 
G^I{}_{J_1 \dots J_{p+2} \, \bar{L}_1 \dots  \bar{L}_p} (\F, {\bar \F})\,
\S^{J_1} \dots \S^{J_{p+2}} \,
{\bar \S}^{ {\bar L}_1 } \dots {\bar \S}^{ {\bar L}_p }~~~~~~ 
\label{ansatz2}
\eea
Here $G^I{}_{J_1 \dots J_{p+2} \, \bar{L}_1 \dots  \bar{L}_p} (\F, {\bar \F})$
are {tensor functions of the K\"ahler metric,
the Riemann curvature $R_{I {\bar 
J} K {\bar L}} \big( \F, \bar{\F} \big) $ and its covariant 
derivatives.} 

The above  representation for $\U^I_2 (\F,  \bar \F, \S , \bar \S)$ follows
from the structure of the transformation laws with respect to 
holomorphic reparametrizations of the K\"ahler manifold $\cM$:
\begin{subequations}
\bea
\F^I  \quad  & \longrightarrow & \quad \F'{}^I=f^I \big( \F \big) ~, \\
\S^I  \quad & \longrightarrow  & \quad \S'{}^I= \frac{\pa f^I (\F) }{\pa \F^J }\S^J~, \\
\U_2^I  \quad &  \longrightarrow &  \quad
\U'{}^I_2 
= \hf  \frac{ \pa^2 f^I  \big( \F \big) }{\pa \F^J \pa \F^K}\, \S^J \S^K
+\frac{ \pa f^I  \big( \F \big) }{\pa \F^J }\, \U_2^J ~.
\eea
\end{subequations}
One can check that
\bea
G^I(\F,  \bar \F, \S , \bar \S) = \frac{1}{6} \nabla_{J_1}R_{J_2 {\bar L} J_3 }{}^I (\F , \bar \F )\,
\S^{J_1}\S^{J_2} \S^{J_3}{\bar \S}^{{\bar L}} +\cO \big( \S^4  \bar \S^2 \big)~.
\eea

Our next step is to require the tangent-bundle action, eq. (\ref{act-tab-mod}), 
to be  supersymmetric.
This proves to imply that  $\cL \big(\F, \bar \F, \S , \bar \S \big)$
and  $G^I(\F,  \bar \F, \S , \bar \S) $  obey the following equations:
\begin{subequations}
\bea
\frac{\pa \cL}{\pa \S^J} \,\frac{\pa G^J}{\pa {\bar \S}^{\bar I} } 
=~ \frac{ \pa \X}{ \pa {\bar \S}^{\bar I} } ~, ~&&
\label{master1} \\
\nabla_I \cL+ \frac{\pa \cL}{\pa \S^J} \,\frac{\pa G^J}{\pa { \S}^{I} } 
= ~\frac{ \pa \X}{ \pa { \S}^{ I} } ~,~&&
\label{master2}  \\
\hf R_{K {\bar I} L }{}^J\, \frac{\pa \cL}{\pa \S^J}\, \S^K \S^L
+ \frac{\pa \cL}{\pa {\bar \S}^{\bar I} }
+g_{J \bar{I}}\, \S^J 
- \frac{\pa \cL}{\pa \S^J} \,
\nabla_{\bar I} G^J =-\nabla_{\bar I} \X~,&&
\label{master3}
\eea
\end{subequations}
where $\X$ turns out to be
\bea
\X = \S^I  \nabla_I \cL + 2 G^I\, \frac{\pa \cL}{\pa \S^I} ~.
\label{X1}
\eea
We  also define
\bea
\nabla_I \cL &:=&\sum_{n=1}^{\infty} 
\Big( \nabla_I \cL_{J_1 \cdots J_n {\bar L}_1 \cdots {\bar 
L}_n }  \big( \F, \bar{\F} \big)\Big)  \S^{J_1} \dots \S^{J_n} 
{\bar \S}^{ {\bar L}_1 } \dots {\bar \S}^{ {\bar L}_n }\non \\
&=& \frac{\pa \cL}{\pa \F^I} - \frac{\pa \cL}{\pa \S^K}\, \G^K_{IJ} \,\S^J~,
\eea
and similarly for $\nabla_{\bar I} G^J$. 

It is natural to analyze how the above relations simplify 
in the special case when $\cM$ is Hermitian symmetric.
It holds that 
\bea
\nabla_L  R_{I_1 {\bar  J}_1 I_2 {\bar J}_2}
= { \nabla}_{\bar L} R_{I_1 {\bar  J}_1 I_2 {\bar J}_2} =0
\quad \Longrightarrow \quad 
\nabla_{ I} \cL = G^I =\X=0 ~.
\eea

In the cotangent-bundle formulation, eq. (\ref{act-ctb}),
 the chiral variables $(\F^I, \J_J)$ are local complex coordinates on  the cotangent 
bundle $T^* \cM$,  and the hyperk\"ahler  potential of $T^* \cM$ is
\bea
{\mathbb K} (\F, \bar \F, \J, \bar \J )
:=K \big( \F, \bar{\F} \big)+  \cH(\F, \bar \F, \J, \bar \J )~.
\label{H-Kpotential}
\eea
Within this formulation, 
the second  supersymmetry can be shown to take the form:
\bea
\d \F^I =\hf {\bar D}^2 \Big\{ \overline{\e \q} \, \frac{\pa \mathbb K}{\pa \J_I} \Big\} ~, 
\qquad
\d \J_I =- \hf {\bar D}^2 \Big\{ \overline{\e \q} \, 
\frac{\pa \mathbb K}{\pa \F^I}   \Big\}~.
\label{SUSY-ctb4}
\eea
Introduce the condensed notation: 
\bea
\f^a := (\F^I\,, \J_I) ~, \qquad {\bar \f}^{\,\bar a} = ({\bar \F}^{\bar I}\,, {\bar \J}_{\bar I}), 
\eea
as well as the  symplectic matrix ${\mathbb J} =({\mathbb J}^{a b} )$, its inverse
${\mathbb J}^{-1} =(-{\mathbb J}_{a b} )$ and their complex conjugates,
\bea
{\mathbb J}^{a b} = {\mathbb J}^{\bar a \bar b} = 
\left(
\begin{array}{rc}
0 ~ &  {\mathbbm 1} \\
-{\mathbbm 1} ~ & 0  
\end{array}
\right)~,  \qquad 
{\mathbb J}_{a b} = {\mathbb J}_{\bar a \bar b} = 
\left( 
\begin{array}{rc}
0 ~ &  {\mathbbm 1} \\
-{\mathbbm 1} ~ & 0  
\end{array}
\right)~. 
\eea
Then the  supersymmetry transformation can be rewritten as 
\bea 
\d \f^a =\hf {\mathbb J}^{ab} \,
{\bar D}^2 \Big\{ \overline{\e \q} \, \frac{\pa \mathbb K}{\pa \f^b} \Big\} 
= \hf {\bar D}^2 \Big\{ \overline{\e \q} \, {\bar {\bm \O}}^a \Big\}
~, \qquad 
 {\bar {\bm \O}}^a := {\mathbb J}^{ab} \, \frac{\pa \mathbb K}{\pa \f^b}~.
\label{SUSY-ctb5}
\eea
These results can now be linked to the general analysis
of $\cN=2$ sigma-models in $\cN=1$ superspace \cite{HKLR}
reviewed in subsection \ref{N=2-->N=1}.  First of all, we see 
that the supersymmetry transformations agrees with the ansatz (\ref{4.16}).

Using eq. (\ref{SUSY-ctb5}), we can read off the expression for 
the holomorphic two-form  on $T^*\cM$.
By definition, the anti-holomorphic two-form is  
\bea
{\bar {\bm \o}}_{\bar b \bar c} = {\bm g}_{a \bar b} \,{\bar {\bm \O}}^a{}_{,\bar c}~,
\non
\eea 
with $ {\bm g}_{a \bar b} $ the K\"ahler metric\footnote{We use a bold-face notation 
for the K\"ahler metric on $T^*\cM$ in order to distinguish it from the metric on $\cM$.}
\bea
{\bm g}_{a \bar b} = \frac{\pa^2 \mathbb K}{  \pa \f^a   \pa {\bar \f}^{\bar b} }
= \left(\begin{array}{cc}
 \frac{\pa^2 \mathbb K}{\pa \F^I \pa {\bar \F}^{\bar J}} 
 ~ &   \frac{\pa^2 \mathbb K}{\pa \F^I \pa {\bar \J}_{\bar J}}  \\
 \frac{\pa^2 \mathbb K}{\pa \J_I \pa {\bar \F}^{\bar J}} 
~ &  \frac{\pa^2 \mathbb K}{\pa \J_I \pa {\bar \J}_{\bar J}} 
\end{array}
\right)~.
\eea
Recalling the explicit form of $ {\bar {\bm \O}}^a $, eq. (\ref{SUSY-ctb5}),
we can immediately see that 
${\bar {\bm \o}}_{\bar b \bar c} $ is indeed  antisymmetric,
\bea
{\bar {\bm \o}}_{\bar a \bar b} 
={\bm g}_{ \bar a c}  \,{\mathbb J}^{cd} \,{\bm g}_{d \bar b}   ~, \qquad
{ {\bm \o}}_{ a b} 
={\bm g}_{  a \bar c}  \,{\mathbb J}^{\bar c \bar d} \,{\bm g}_{\bar d  b}   ~
 \eea
 
Direct calculations show that  
\bea
{\bm \o}_{ab} = {\mathbb J}_{ab} + \cO (\J \bar \J )~. 
\eea
Since ${\bm \o}_{ab}$ must be holomorphic, we  immediately conclude that
\bea
{\bm \o}_{ab} = {\mathbb J}_{ab}~, \qquad
{\bar {\bm \o}}_{\bar a \bar b} = {\mathbb J}_{\bar a \bar b}\quad \Longrightarrow \quad
{ {\bm \o}}^{ a  b} = {\bm g}^{a \bar  c}  {\bm g}^{b \bar  d} {\bar {\bm \o}}_{\bar c \bar d} 
= {\mathbb J}^{ab}~.
\eea
As a result, the holomorphic symplectic two-form
${\bm \o}^{(2,0)}$ of  $T^*\cM$ coincides with 
the canonical holomorphic symplectic two-form, 
\bea
{\bm \o}^{(2,0)} := \hf {\bm \o}_{ab}\, {\rm d}\f^a \wedge {\rm d} \f^b
 =  {\rm d} \F^I \wedge{\rm d}\J_I~.
\eea

\section{Topics not covered} 
\setcounter{equation}{0}

These lectures, which reflect the author's interests,  
have not touched upon several important topics concerning sigma-models 
in projective superspace. Here we would like to make a few comments about some of these 
developments.

There exists a large body of research literature on  sigma-model couplings of $\cN=2$ 
tensor multiplets, including the pioneering papers \cite{HitchinKLR,KLR,LR}.
A complete list of references can be found in \cite{LR2008}. Self-couplings of $\cO(2n)$ multiplets, 
with $n\geq 2$, are less studied, see \cite{LR2008} for a review.

Off-shell 4D $\cN=2$ superconformal multiplets in projective superspace and their general sigma-model
couplings were presented  in \cite{K-hyper}. In the case of $\cN=2$ tensor multiplets, their
superconformal couplings were described much earlier in \cite{KLR,BS,deWRV}.
The most general $\cN=2$ superconformal sigma-models can be realized in terms of 
polar multiplets \cite{K-hyper}. They are described by the action (\ref{nact}) in which 
 the K\"ahler potential obeys the homogeneity condition\footnote{The action (\ref{nact4})
with the K\"ahler potential obeying  the homogeneity condition (\ref{Kkahler2}) defines a general 
$\cN=1$ superconformal sigma-model, see \cite{K09} for more details.}
\bea
\F^I \frac{\pa}{\pa \F^I} K(\F, \bar \F) =  K( \F,   \bar \F)~.
\label{Kkahler2}
\eea
The geometric interpretation of such sigma-models, 
albeit realized in a somewhat different form, was given in \cite{KLvU}.
Their formulation in terms of $\cN=1$ chiral superfields, which is obtained
upon elimination of the polar multiplet auxiliaries, was presented in \cite{K09}.

The projective superspace approach was extended to six \cite{GrunL,GPT-M} and five
\cite{KL,K-compactified} dimensions (with Ref. \cite{K-compactified} devoted to 
5D off-shell superconformal sigma-models).

General off-shell $\cN=2$ locally supersymmetric nonlinear sigma-models were constructed in 
\cite{KLRT-M}. Their properties remain largely unexplored.

Interesting geometric aspects of sigma-models in projective superspace were uncovered in \cite{LR2008}.
\\

\noindent
{\bf Acknowledgements:}
The author is happy to thank the organizers of the 30th Winter School ``Geometry and Physics'' at Srni, 
particularly Vladimir Sou\v{c}ek and Rikard von Unge, for the opportunity to participate in the school.
Discussions with Ulf Lindstr\"om and Rikard von Unge are gratefully acknowledged. 
The author also thanks Ian McArthur,  Joseph Novak and Rikard von Unge
for helpful comments on the manuscript, 
and Jim Gates for conversations about the history of off-shell $\cN=1$ scalar multiplets and duality.
This work  is supported in part by the Australian Research Council and the Australian Academy of Science.

\appendix

\section{Superconformal group} 
\setcounter{equation}{0}
This appendix contains a summary of the 
4D $\cN$-extended superconformal group $ {\rm SU}(2,2|\cN)$.
Any element $g \in  {\rm SU}(2,2|\cN)$ is a $(4+\cN)\times (4+\cN)$ 
{supermatrix} of the form:
\bea
g^\dagger \O g = \O, \qquad {\rm Ber}\, g=1~, 
\qquad
\O =\left(
\begin{array}{c | c ||c}
0 ~ & ~  \mathbbm{1}_2&{ 0}   \\
\hline
 \mathbbm{1}_2  & ~0&0 \\
\hline
\hline
0 ~& ~{ 0}&-\mathbbm{1}_\cN
\end{array}
\right)~,
\eea
where ${\rm Ber}\, A$ stands for the superdeterminant of a supermatrix $A$ \cite{Berezin,Berezin2}.
In a neighborhood of the unit element of $ {\rm SU}(2,2|\cN)$, every group element can be 
represented in the exponential form:
\bea 
{g} ={\rm e}^{L}~, \qquad L^\dagger \O  +\O L  =0~, \qquad {\rm str}\, L =0~,
\eea
with $L$  an element of the superconformal algebra, $ {\rm su}(2,2|\cN)$. 
The most general expression for $L$ is as follows:
\bea
{ L} = \left(
\begin{array}{c |c ||c}
\o_\a{}^\b - \s \d_\a{}^\b  \quad &  -{\rm i} \,a_{\a \dt \b} \quad &
2\eta_\a{}^j \\
\hline
\phantom{\Big|} -{\rm i} \,b^{\dt \a \b} \quad & -{\bar \o}^{\dt \a}{}_{\dt \b} 
+ {\bar \s}  \d^{\dt \a}{}_{\dt \b}   \quad &
2{\bar \e}^{\dt \a j} \\
\hline 
\hline
\phantom{\Big|} 2\e_i{}^\b \quad & 2{\bar \eta}_{i \dt \b} \quad & \frac{2}{\cN}({\bar \s} - \s)\,
\d_i{}^j +  \L_i{}^j
\end{array}
\right)~,~~~
\label{su(2,2|n)}
\eea
where
\be
\s = \hf \left( \t + {\rm i}\, 
\frac{\cN}{\cN -4} \vf \right)~,
\qquad
\L^\dag =  -\L~, \qquad  {\rm tr}\; \L = 0~.
\ee
Here the matrix elements, which 
are not associated with the super-Poincar\'e transformations (\ref{SP2}) and (\ref{SP-lor}),  
correspond to a special conformal transformation ${\bm a}=(a_{\a \dt \b} ) ={\bm a}^\dagger$,
$S$--supersymmetry $(\eta_\a^i,~{\bar \eta}_{i \dt \a})$,
combined scale and chiral transformation $\s$, 
and chiral $SU(\cN)$ transformation $\L_i{}^j$. The case $\cN=4$ requires a special consideration.

The 4D $\cN=1$ superconformal group was introduced by Wess and Zumino \cite{WZ}.

\section{Canonical coordinates for K\"ahler manifolds}
\setcounter{equation}{0}

In this appendix, we recall the concept of canonical coordinates for K\"ahler manifolds
\cite{Bochner}.

Given a K\"ahler manifold $\cM$,  
for any point $ p_0 \in \cM$ there exists  a  neighborhood of $p_0$ such that 
holomorphic reparametrizations  and K\"ahler transformations
can be used to choose  coordinates with origin at $p_0$ in which
the K\"ahler potential is
\bea
{K} (\F, \bar \F ) &=&{g}_{I \bar{J}}| \,\F^I {\bar \F}^{\bar J}
+ \sum^{\infty}_{ m,n \geq 2}  
{ { K}^{(m,n)} (\F, \bar \F)}~,
\non \\
 { K}^{(m,n)} (\F, \bar \F) &:=&
\frac{1}{m! n!}\,
{ K}_{I_1 \cdots I_m {\bar J}_1 \cdots {\bar J}_n }|  \, \F^{I_1} \dots \F^{I_m} 
{\bar \F}^{ {\bar J}_1 } \dots {\bar \F}^{ {\bar J}_n }~.
\label{normal-gauge} 
\eea
Such a coordinate system in the K\"ahler manifold is called {\it canonical}. 
It was first introduced by Bochner \cite{Bochner} and extensively used by Calabi \cite{Calabi}.
There still remains  freedom to perform 
linear reparametrizations which can be used 
to set the metric at the origin, $p \in \cM$, to  be ${ g}_{I \bar{J}}|= \d_{I \bar{J}}$.
It turns out that the  coefficients ${ K}_{I_1 \cdots I_m {\bar J}_1 \cdots {\bar J}_n } |$
are tensor functions of the K\"ahler metric ${g}_{I \bar{J}}|$,
the Riemann curvature $R_{I {\bar J} K {\bar L}}  |$ and its covariant 
derivatives,    evaluated at the origin.
In particular, one finds
\begin{subequations}
\bea
{K}^{(2,2)} &=& \frac{1}{4} R_{I_1 {\bar J}_1 I_2 {\bar J}_2} |\, \F^{I_1}\F^{I_2}
{\bar \F}^{{\bar J}_1}{\bar \F}^{{\bar J}_2}~,
\label{(2,2)}  \\
 {K}^{(3,2)} &=& \frac{1}{12} 
\nabla_{I_3} R_{I_1 {\bar J}_1 I_2 {\bar J}_2}| \,
\F^{I_1} \dots \F^{I_3}
{\bar \F}^{{\bar J}_1}{\bar \F}^{{\bar J}_2}~, 
\label{(3,2)} \\
 {K}^{(4,2)} &=& \frac{1}{48} 
\nabla_{I_3} \nabla_{I_4} R_{I_1 {\bar J}_1 I_2 {\bar J}_2} |\,
\F^{I_1} \dots \F^{I_4}
{\bar \F}^{{\bar J}_1}{\bar \F}^{{\bar J}_2}~, 
\label{(4,2) } \\
{K}^{(3,3)} &=& \frac{1}{12} \Big\{ \frac{1}{6} 
\{ \nabla_{I_3}, {\bar \nabla}_{{\bar J}_3} \}
R_{I_1 {\bar J}_1 I_2 {\bar J}_2} |
+R_{I_1 {\bar J}_1 I_2 }{}^L |R_{L {\bar J}_2 I_3 {\bar J}_3}| \Big\}  \non \\
 &&\quad  \times \F^{I_1} \dots \F^{I_3}
{\bar \F}^{{\bar J}_1}\dots{\bar \F}^{{\bar J}_3}~~~~~~~
\label{(3,3)}  
\eea
\end{subequations}
In the  modern literature on supersymmetric sigma-models, some authors, 
being unaware of the work of \cite{Bochner},
refer to the canonical coordinates introduced  as a 
{\it normal gauge} \cite{GGRS} or {\it K\"ahler normal coordinates} \cite{HIN}.

If $\cM$ is Hermitian symmetric, then  
\bea
\nabla_L  R_{I_1 {\bar  J}_1 I_2 {\bar J}_2}
= {\bar \nabla}_{\bar L} R_{I_1 {\bar  J}_1 I_2 {\bar J}_2} =0
\qquad \Longrightarrow \qquad 
{ K}^{(m,n)}=0~, \quad m\neq n~.
\label{covar-const}
\eea
This follows from the fact that, 
for Hermitian symmetric spaces, there exists a closed-form expression for
the K\"ahler potential in the canonical coordinates \cite{KN}:
\bea
K \big( \F, \bar \F   \big) = - \hf {\bm \F}^{\rm T} {\bm g} \,
\frac{ \ln \big( {\mathbbm 1} - {\bm R}_{\F,\bar \F}\big)}{\bm R_{\F,\bar \F}}
\, {\bm \F}~, \qquad 
{\bm \F} :=\left(
\begin{array}{c}
\F^I\\
{\bar \F}^{\bar I} 
\end{array}
\right) ~.
\eea
Here we have introduced
\bea
{\bm g}
&:=&\left(
\begin{array}{cc}
0 & g_{I \bar J}|\\
g_{{\bar I}J} |&0 
\end{array}
\right)
~, \qquad
{\bm R}_{\F,\bar \F}
:= \left(
\begin{array}{cc}
0 & (R_\F)^I{}_{\bar J}\\
(R_{\bar \F})^{\bar I}{}_J &0 
\end{array}
\right)~,  \non \\
(R_\F)^I{}_{\bar J} &:=&\hf R_K{}^I{}_{L \bar J}|\, \F^K \F^L~,  
\qquad (R_{\bar \F})^{\bar I}{}_J := \overline{(R_\F)^I{}_{\bar J}}~.
\eea

\section{Two-component (iso)spinor conventions}
\setcounter{equation}{0}
In the case of two-component undotted spinors, such as $\j_\a $ and $\j^\a$, their indices are raised and lowered by the rule:
\bea
 \j^\a =  \ve^{\a \b} \,\j_\b~, \qquad \j_\a =  \ve_{\a \b} \,\j^\b ~,
\eea
where $ \ve^{\a \b} $ and $ { \ve_{\a \b}} $ are $2\times 2 $ 
antisymmetric matrices normalized as
\bea
\ve^{12}= \ve_{21} =1~.
\label{c2}
\eea
The same conventions are used for dotted spinors (${\bar  \j}_{\dt \a} $ and ${\bar  \j}^{\dt \a}$), 
and for SU(2) isospinors ($ v_i $ and $ v^i$), in particular
\bea
v^i =\ve^{ij} \, v_j ~, \qquad v_i= \ve_{ij} \,v^j~.
\eea
The SL$(2,{\mathbb C})$ invariant antisymmetric tensors $\ve^{{\dt \a} \dt \b} $
and $\ve_{{\dt \a} \dt \b} $ and the SU(2) invariant antisymmetric tensors
$\ve^{ij}  $ and $\ve_{ij} $ are normalized as in (\ref{c2}).

Lorentz-invariant spinor
bi-products  are defined by 
\be
\j \l = \j^\a \l_\a ~, \quad \j^2 = \j \j~, \quad 
\qquad {\bar \j} \bar \l = {\bar  \j}_{\dt \a} {\bar  \l}^{\dt \a} ~, \quad {\bar \j}^2 = {\bar \j} {\bar \j} ~.
\ee

\small{

}

\end{document}